\documentclass[preprint,showpacs,preprintnumbers,amsmath,amssymb]{revtex4}

\usepackage{graphicx}
\usepackage{dcolumn}
\usepackage{bm}
\usepackage{times}

\def\refs#1#2#3#4#5#6{#1, {#3} {\bf #4}, #5(#6)}
\def\nuc#1#2{$^{#1}$#2}

\begin{document}
\title{
Precise measurement of cross section of $^3$He($^3$He,2p)$^4$He by using 
He-3 doubly charged beam
}

\author{
Nobuyuki Kudomi$^{1}$, 
Masataka Komori$^{2}$, 
Keiji Takahisa$^{1}$, 
Sei Yoshida$^{1}$, 
Kyo Kume$^{3}$, 
Hideaki Ohsumi$^{4}$, and  
Takahisa Itahashi$^{1}$ 
}
\email{kudomi@rcnp.osaka-u.ac.jp}
\affiliation{
$^{1}$
Research Center for Nuclear Physics, 
Osaka University, Ibaraki, Osaka 567-0047, Japan \\
$^{2}$
National Institute of Radiological sciences 4-9-1 Anagawa Inageku Chibashi, 
Japan 263-8555 \\
$^{3}$
The Wakasa Wan Energy Research Center 64-52-1 Nagatani, Tsuruga, 
Fukui Japan 914-0192 \\
$^{4}$
Fuculty of Culture and Education, Saga University, 1 Honjouchou Sagashi, 
Japan 840-0027
}

\begin{abstract}
The fusion cross section of \nuc{3}{He}(\nuc{3}{He},2p)\nuc{4}{He} 
at a center of mass energy of 30 to 50 keV 
has been measured 
by using helium-3 doubly ionized beam 
at a low-energy high current accelerator facility, OCEAN. 
Free from molecular interference in the beam, the measurement determines 
the astrophysical S-factor with better statistical and systematical errors 
than previous data. By using singly and doubly charged helium-3 ions 
the facility envisages to provide the data from high energy to Gamow energy regions. 
\end{abstract}


\maketitle

\section{Introduction}
Of the reactions involved the solar combustion, namely 
d + p $\rightarrow$ \nuc{3}{He} + $\gamma$, 
\nuc{3}{He} + \nuc{3}{He} $\rightarrow$ 2p + $\alpha$, 
and \nuc{3}{He} + $\alpha$ $\rightarrow$ \nuc{7}{Be} + $\gamma$, 
we have focused on the cross section 
measurement of the \nuc{3}{He} + \nuc{3}{He} reaction 
at an effective energy of E$_{\rm cm}$=30-50 keV.  
Among many experimental works, 
the only experiment which has been conducted in or lower than the Gamow energy 
is the \nuc{3}{He} + \nuc{3}{He} $\rightarrow$ 2p + $\alpha$ reaction 
at the LUNA in Gran Sasso laboratory \cite{LUNA}. 
The latter was performed near the Gamow energy 
without measurement at  E$_{\rm cm}$=25-45 keV 
due to accelerator limitations 
that lower voltage could not be applied with a high voltage static accelerator 
while higher voltage more than 50 kV could be impossible at the LUNA. 
For a wider energy range E$_{\rm cm}$=17.9 to 342.5 keV, 
the experiments done by Krauss et al 
slightly extended the data over this energy gap \cite{LUNA2}.  
For the nuclear astrophysics discussion, 
in particular, 
standard solar model or nucleo-systhesis 
continuous data down to low energy 
is crucial to deduce the astrophysical S-factor, S$_{17}$. 
Therefore, successive and precise data 
from around 50 keV to 20 keV center of mass energy are needed. 
For this purpose, 
we have constructed a low energy and compact accelerator facility 
which provides doubly charged \nuc{3}{He} ions for the measurement 
in the region between 50 keV to 25 keV and also singly charged \nuc{3}{He} ions 
in the region less than 25 keV center of mass energy. 
This is the first report of a series experiments done at our 
low-energy high current accelerator facility, OCEAN. 
We obtained precise results from 45.3 to 31.2 keV 
in center of mass energy for the \nuc{3}{He}+\nuc{3}{He}$\rightarrow$2p+$\alpha$ 
reaction 
with a doubly charged incident \nuc{3}{He} beam. 
It has the substantial advantage of eliminating the molecular interference 
in the proton spectra by the \nuc{3}{He} + d reaction caused 
by the incident HD$^{+}$ beam. 

\section{Experimental apparatus}
\begin{figure}
\includegraphics*[width=150mm]{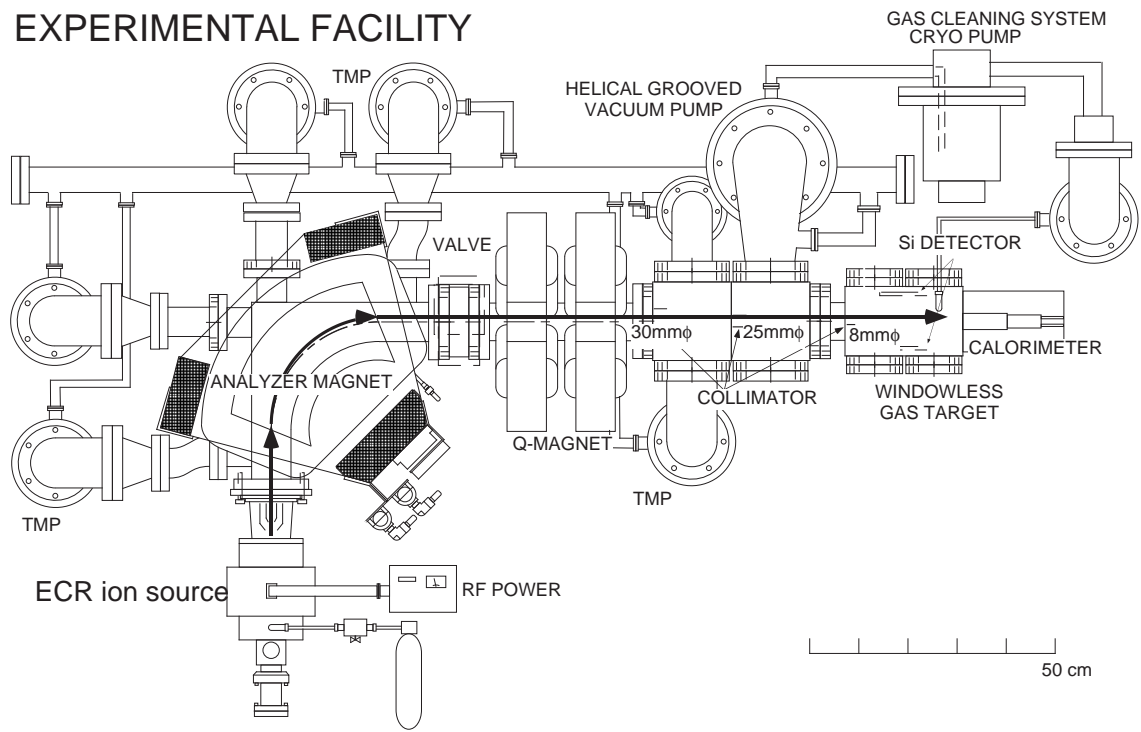}
\caption{
Complete layout of {\underline O}saka University 
{\underline C}osmological 
{\underline E}xperimental 
{\underline A}pparatus for 
{\underline N}uclear Physics, 
OCEAN .  
}
\label{fig:layout}
\end{figure}

The experimental apparatus, 
OCEAN (Osaka University Cosmological Experimental Apparatus for Nuclear Physics) 
consists of: 
(1) a powerful ion source that provides an intense current of more than 1 mA 
for \nuc{3}{He}$^{1+}$ ions at incident energies of 30-50 keV, 
or 200 $\mu$A for \nuc{3}{He}$^{2+}$ at incident energies of 90.6 to 62.4 keV. 
(2) low energy beam transport with good transmission, 
(3) a windowless gas target and recirculation / purification system, 
(4) a reliable calorimeter, 
(5) detectors for reaction identification, and 
(6) an electronics and data acquisition system based on CAMAC. 
The layout of the OCEAN is shown in Fig. \ref{fig:layout}. 
  
\subsection{Ion source and extraction electrodes}
An intense ion source that can produce \nuc{3}{He}$^{2+}$ ions 
is essential for the present study. 
The NANOGUN$^{\rm TM}$, which was obtained from PANTECH, 
confines high-temperature electrons produced by the 
electron cyclotron resonance (ECR) and is assembled 
into an ECR ion source with 10 GHz, 
200 watt RF generator (model VZX-6383G5, CPI). 
The original NANOGUN ECR ion source made by PANTECH can 
provide 40 $\mu$A for \nuc{40}{Ar}$^{8+}$ at 20 kV extraction voltage 
with an RF power of 60 W. 
From these data we could easily foresee the possibilities of obtaining 
\nuc{3}{He}$^{1+}$ or \nuc{3}{He}$^{2+}$ ion beams of more than 
100 $\mu$A, although an applied potential to the extraction 
is not enough to use this ion source for
astrophysical applications in a wider enery range. 
For this purpose, 
we redesigned the ion source extraction system to meet the ion optical condition 
for the present window-less gas target. 
Leroy et al. reported an improvement for the analyzed current and beam emittance 
of \nuc{3}{He}$^{1+}$ beam by 
the a system called Multielectrodes Extraction 
for CAPRICE-type ECR ion source.  
It supplied with an analysed current of 2.3 mA of \nuc{3}{He}$^{1+}$ 
giving a transmission of more than 75 \% in the beam line \cite{Leroy}.  
 did you also do this). 
Based on several experimental studies and computer simulations 
we designed and applied a two-electrode extraction system for 
the NANOGUN ECR ion source \cite{Ita_ION}. 
This improves the beam emittance under the influence of a strong space charge force, 
and secondly, 
it moderates the electric field gradient ascribed to high operational voltage. 
The optimization process for applying a suitable voltage to the intermediate electrode 
progressed considerably during experiments measured at each center of mass energy. 
For the present experiment between 45 keV and 31.2 keV center of mass energy, 
around 100 $\mu$A of double charged \nuc{3}{He} beam could pass 
through the three stages collimators to the gas target. 
A specially designed ceramic insulator (SUMIKIN CERAMIC Co-Ltd) 
with two folds on the surface could support up to 50 kV potential 
without any surface discharge at a distance of 170 mm from the surface. 
We fabricated conic type and straight type extraction electrodes. 
It consists of a beam forming electrode which has a long straight part similar to the 
original NANOGUN's design and a conic structure for the beam forming electrode 
nearly parallel to the end surface of the plasma chamber. 
The performance and design have been described in previous papers \cite{Ita_ION}. 

\subsection{Low energy beam transport}
\begin{figure}[hbtp]
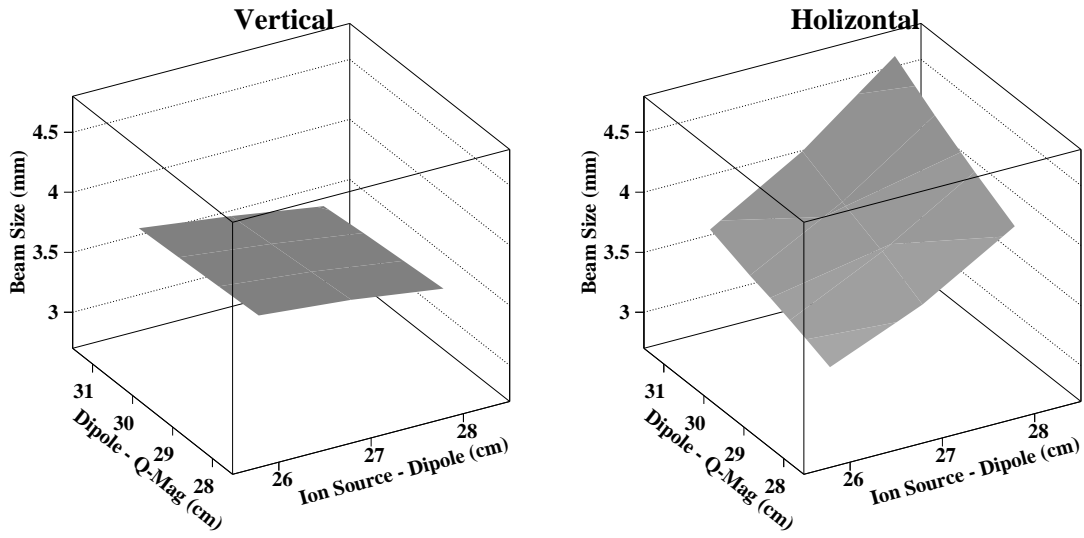

\includegraphics*[width=75mm]{gios-y.epsf}
\includegraphics*[width=75mm]{gios-x.epsf}
\caption[gios]{
Beam transport calculation for the present designed scheme consisting of a 
90 $^{\circ}$ dipole magnet+quadrupole doublet 
for 1 mA 50 keV \nuc{3}{He}$^{1+}$ beam. 
The beam size at the target (vertical and horizontal) was calculated 
as a function of the distance between the ion source to the dipole magnet 
or dipole magnet to Q-magnet, respectively. 
The beam source is assumed to be 100 $\pi$mm mrad and 5 mm in diameter. 
}
\label{fig:gios}
\end{figure}

The low energy beam transport system between the ECR ion source 
and the gas target 
achieved a high beam intensity and other desirable beam qualities, 
thereby allowing for precision measurements of the present 
\nuc{3}{He}+\nuc{3}{He}$\rightarrow$2p+$\alpha$ experiment. 
Generally, 
it is known that there a strong space charge effect 
in the beam transport at ion currents of more than 1 mA. 
It is essential that this effect is accounted for 
when calculating the beam optics. 
We used a GIOS code developed by Wollnick et al.  
for incorporating this effect \cite{wollnic}. 
We adopted the D (dipole, 90-degree deflection angle) + Q + Q transport scheme 
for our system since it is easier to operate fewer elements. 
In this calculation 
we assumed a beam source of 100 $\pi$ mm mrad and 5mm in diameter. 
Despite the variance in total potential of the beam, 
a nearly invariant beam form could be realized 
at the source exit using the extraction system stated before. 
To maintain the minimum slit aperture, 
we calculated the dimension of the beam at the target position 
by varying the various parameters of elements and drift lengths 
so as to achieve smaller $dx$ and $dy$. 
Very attractive results such as a constant $dx$ and $dy$ 
and a nearly parallel beam form are shown in Fig. \ref{fig:gios}. 
The beam transmission efficiency from the ion source 
through the target is about 30 \% (Table \ref{tab:is2target}). 

\subsection{Window-less gas target}
\begin{figure}
\includegraphics*[width=150mm]{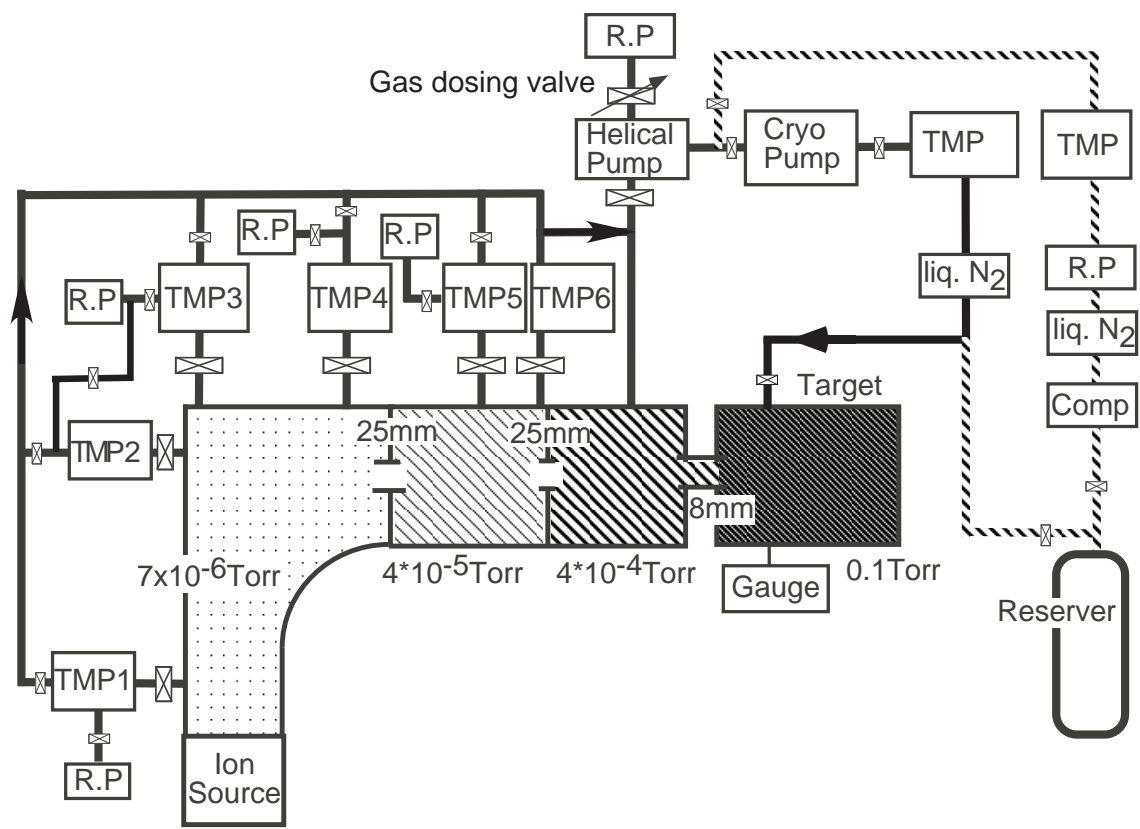}
\caption{
Complete layout of the window-less gas target evacuation and 
recirculation system.  
TMP: Turbo molecular pump.  
R.P: Rotary pump.  
}
\label{fig:circulation}
\end{figure}

The window-less gas target for the study of the \nuc{3}{He}+\nuc{3}{He} reaction 
consists of  a differential pumping and gas circulation/purification system 
as shown in Fig. \ref{fig:circulation}. 

In order to maintain a pressure of 1 mbar in the chamber of 
the \nuc{3}{He} gas target without window.  
the pumping system should be composed of several stages 
between the target chamber and beam transport system. 
Thus we prepared a helical groove vacuum pump 
(model TS-440, OSAKA VACUUM Co.Ltd TS-440) 
as the main pump for evacuating the gas flow at the viscose region 
as well as at the higher vacuum region. 
The sizes of collimators at every stage were estimated 
by calculation in order to maintain a pressure 
of about 0.3 Torr in the target chamber. 
During the experiment, the pressure in the target chamber
was maintained at 3.1($\pm$0.1) Torr for about four days. 
 
The recirculation system consists of a helium tight pump, 
an oil free diaphragm membrane compressor, 
reserver vessels, 
compound gauges, 
ultra fine regulated valves and a quadruple mass spectrometer 
as shown in Fig. \ref{fig:circulation}. 
Since all evacuation pumps connected to the helical pump, 
a gas dosing system has been employed in order to maintain a constant 
target pressure to evacuate the input gas amount at the ion source.  
Because during experiments, gas injection into the ECR ion source 
had the effect of gradually increasing the target pressure. 
Thus, a constant target pressure was maintained 
by controlling the signal from the capacitance manometer (Barocel-655) 
located at the target chamber 
with a gas dosing apparatus (model EVR 116 and RVC 300 controller).  

The new purification system developed for the present experiment 
is quite different from the usual method. 
It exploits a cryo-pump (model U-140W, DAIKIN)
as a purifier without a special absorbent and liquid nitrogen, 
although high heat input could be expected at the high operating pressure. 
This has been overcome by adding another oil free turbo-molecular pump 
(model A30FC, ANELVA) 
between the cryo-pump and the target chamber. 
As pointed out by  A. Krauss et al. \cite{LUNA2}, 
the deuterium contamination both in the target and the beam, resulting 
from the water vapor, is a crucial problem 
for obtaining low energy data, since the d + \nuc{3}{He} reaction cross section 
is six orders of magnitude higher than that of the \nuc{3}{He} + \nuc{3}{He} reaction. 
In the case of \nuc{3}{He}$^{2+}$ beam (e/m=2/3), 
we can avoid molecular 
interference of the HD$^{+}$ beam (e/m=1/3), 
{
with analyzer magnet (Fig. \ref{fig:layout})}
contaminating the incident \nuc{3}{He}$^{1+}$ (e/m=1/3) 
beam, whereas this could not be avoided if we had 
employed a single charged \nuc{3}{He} beam (e/m=1/3),  
we have measured the deuterium contamination in commercial \nuc{3}{He} gas 
by detecting HD$^{+}$ separately via Accelerator Mass Spectrometry (AMS). 
The experiment was carried out using the RCNP K=140 AVF cyclotron. 
The cyclotron accelerator and the NEOMAFIOS ECR (NEOMAFIOS) ion source 
were operated only for the experiment on the beam injection line 
at an anex leading to the 
post accelerator (RCNP Ring cyclotron). 
The present result is 
HD$^{+}$/\nuc{3}{He} = (3.82$\pm$0.69)$\times$10$^{-5}$.  
It is noticed that even if an ECR ion source uses electrons with higher energy 
than that of a duo-plasmatron ion source 
there is a considerable amount of HD$^{+}$ production 
from the surface of the ion source and 
from the bottle of \nuc{3}{He} gas. 

Deuteron contamination in the target was also estimated during operation 
of the recirculation and purification system. 
Then the pressure at the target chamber was 1.2$\times$10$^{-2}$ Torr, 
and 7.6$\times$10$^{-7}$ torr at the helical pump. 
The H$_2$O component 
in the residual gas was measured by means of a quadrupole mass spectrometer, 
to be about 20 \%. 
Assuming that the amount of H$_2$O at the target gas 
is the same as the residual gas,  
and the deuteron abundance is the same as the natural abundance (0.014 \%), 
we can deduce that the deuteron contamination (D$_2$O) is in the order of ppm. 
This is satisfactory for the present measurement as will discussed later. 
In addition, the amount of deuterium contamination in the target gas was also 
evaluated by means of the \nuc{3}{He} + \nuc{3}{He} experiment, 
by detecting the 14.7 MeV proton, and was found to be about 0.1 ppm. 

\subsection{Calorimeter}
A calorimetric device has been developed 
to measure the projectile flux in the present experiment, 
since the conventional method of charge integrating to determine 
the number of incident particles 
is not applicable due to the neutralization of the incident charged particles 
with the gaseous target. 

Many types of calorimeters have been developed \cite{calori}. 
There are two types of calorimeters. 
One type measures the temperature difference between two parts thermally 
isolated with large heat resistance material. 
The other type measures the power needed to stabilize 
the temperature of water 
used to cool the calorimeter cup 
which is bombarded with energetic particles. 
 
We developed the calorimeter by using a heat flux sensor 
with an accuracy of better than 5 \% in 
the dynamical range of 1-30 watts \cite{calori_kudomi}. 
 
\begin{figure}
\includegraphics*[width=150mm]{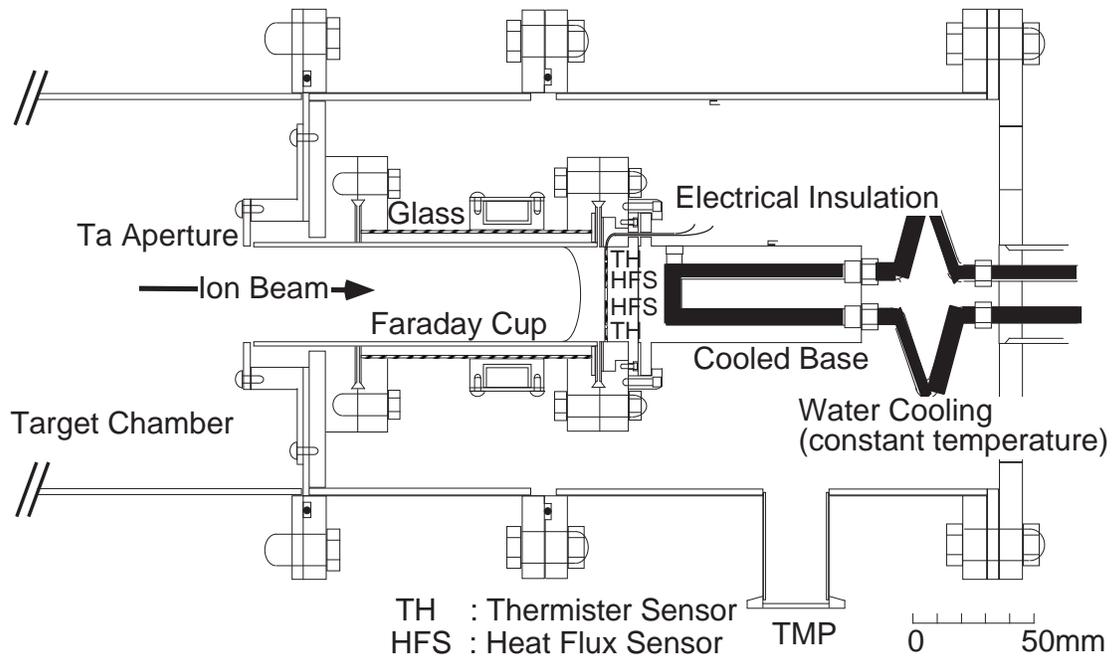}
\caption{
Cross sectional view of the calorimeter.}
\label{fig:calorimeter}
\end{figure}

The structure of calorimeter is shown in Fig. \ref{fig:calorimeter}. 
It consists of a solid copper heat sink (100 mm length, 38 mm diameter) 
with water channels, 
and a Faraday cup (140 mm length, 38 mm diameter, 1.5mm wall thickness) 
in front of the heat sink. 
The Faraday cup is supported by a Pyrex glass insulator 
as well as a metal flange with a Teflon gasket. 
This organic gasket served as a vacuum seal as well as 
an electric insulator. 
Therefore, this calorimeter can also measure the number of incident particles,
when the target chamber is in vacuum, 
by an electrical method.
At the entrance of the cup, 
a secondary electron suppressor is installed. 
Around 100 volt was applied to the suppressor for the current measurement. 

After passing through the window-less gas target, 
the incident \nuc{3}{He}$^{1+}$ and \nuc{3}{He}$^{2+}$ beam is 
neutralized and captured in a Faraday cup and calorimeter. 
In order to be applicable to a wide current range, 
a heat flux sensor (HFS) (OMEGA HFS-3) was used 
to measure the heat transfer from the hot part to the cold part. 
The HFS is a thermister bolometer 
which can measure the heat flux to or from a surface 
with minimum disturbance of the exiting heat flow. 
In this method, a self-generating thermopile is arranged 
around a thin thermal barrier to produce a voltage that is 
a function of the thermal energy passing through the sensor. 
The response of the sensor to the thermal energy input is 1.10 to 1.11 mV/W/m$^2$. 
For precise measurements of the heat transfer, 
we made use of the following procedures: 
1) In vacuum, 
the current of the ion beam can be measured by standard charge integration 
where the calorimeter cup serves as a Faraday cup. 
2) To reduce the conduction and convection losses, 
the heat capacity of the calorimeter 
should be reduced by as much as possible in order to get a better time response. 
With the present heat flux sensor, 
temperature differences of less than 0.001 $^{\circ}$C can be detected easily. 
Therefore, the temperature of the heat sink of the calorimeter 
should be stabilized to better than 0.1 $^{\circ}$C with heat exchange. 
As shown in Fig. \ref{fig:calorimeter}, 
between the copper base and the thin plate of the calorimeter, 
two heat flux sensors are sandwiched with thermister temperature sensors. 
These are originally insulated electrically. 
As shown in Fig. \ref{fig:calorimeter}, 
termister temperature sensors are also located 
to measure the temperature of the ambient or Faraday cup base and the cooled heat sink. 
These are installed in a stainless steel pipe 
(40 mm length, 10.5 cm diameter), 
which can be evacuated by a small turbo-molecular pump. 
 
The calorimeter was tested by using a \nuc{3}{He}$^{2+}$ beam of 
{
energy} of 40 kV. 
The relation between beam current(I) and heat flux(H) can be written as, 
\begin{equation}
I \cdot \delta t = k_1 \cdot H \cdot \delta t  + C \cdot \delta T ,
\label{eq:a}
\end{equation}
where $T$ is the temperature of the calorimeter and $C$ is the heat capacitance. 
The term $C \cdot \delta T$ shows that the temperature of the calorimeter 
depends on the incident beam current. 
Thus, if the intensity I is changed, 
the converted heat is used to heat the calorimeter 
and is also transferred to the cold base. 
 
{
Since the transferred heat may approximated to be proportinal 
to the temperature difference between the the front and cold base,  
the second term }
{
in equation \ref{eq:a}} $C \cdot \delta T $ 
can be rewritten as $k_2 \cdot \delta H $. 
Thus, equation (1) can be written as, 
\begin{equation}
I = k_1 \cdot H  + k_2 \frac{\delta H}{\delta t}.
\end{equation}

\begin{figure}
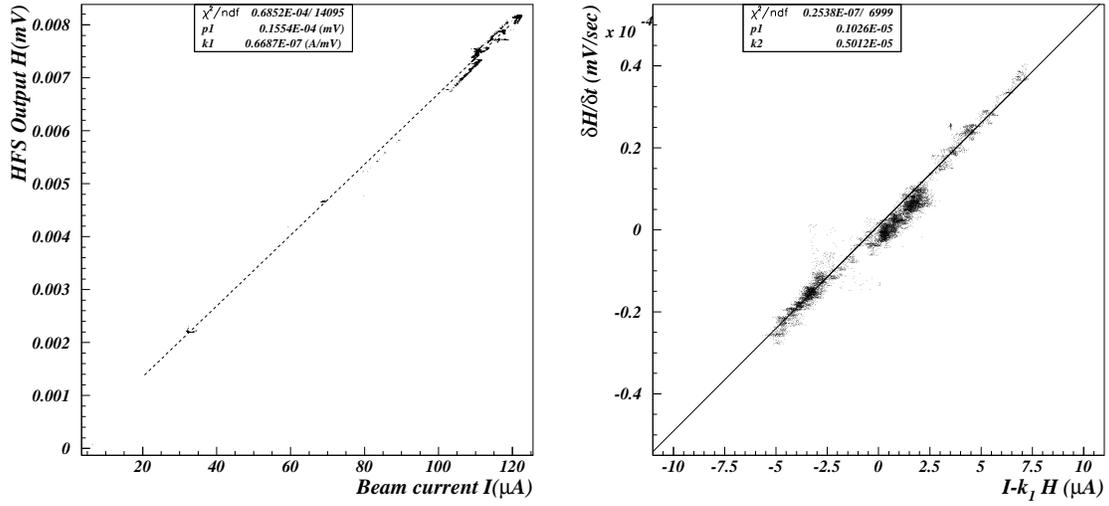

\includegraphics*[width=75mm]{k1.epsf}
\includegraphics*[width=75mm]{k2.epsf}
\caption{
(a)Measured beam current($I (\mu$A)) v.s. 
heat flux($H$ (mV)), under the condition of 
$\delta H/\delta t=0$ in eq. (2). 
The parameter $k_1$ in eq. (2) 
is derived by fitting with a linear function. 
(b)Plot of $\delta H/\delta t$(eq. (2) in text) as 
the difference of 
$I-k_1 \cdot H$. 
The parameter $k_2$ in eq. (2) 
is derived by fitting with a linear function. 
}
\label{fig:k2}
\end{figure}

\begin{figure}
\includegraphics*[width=150mm]{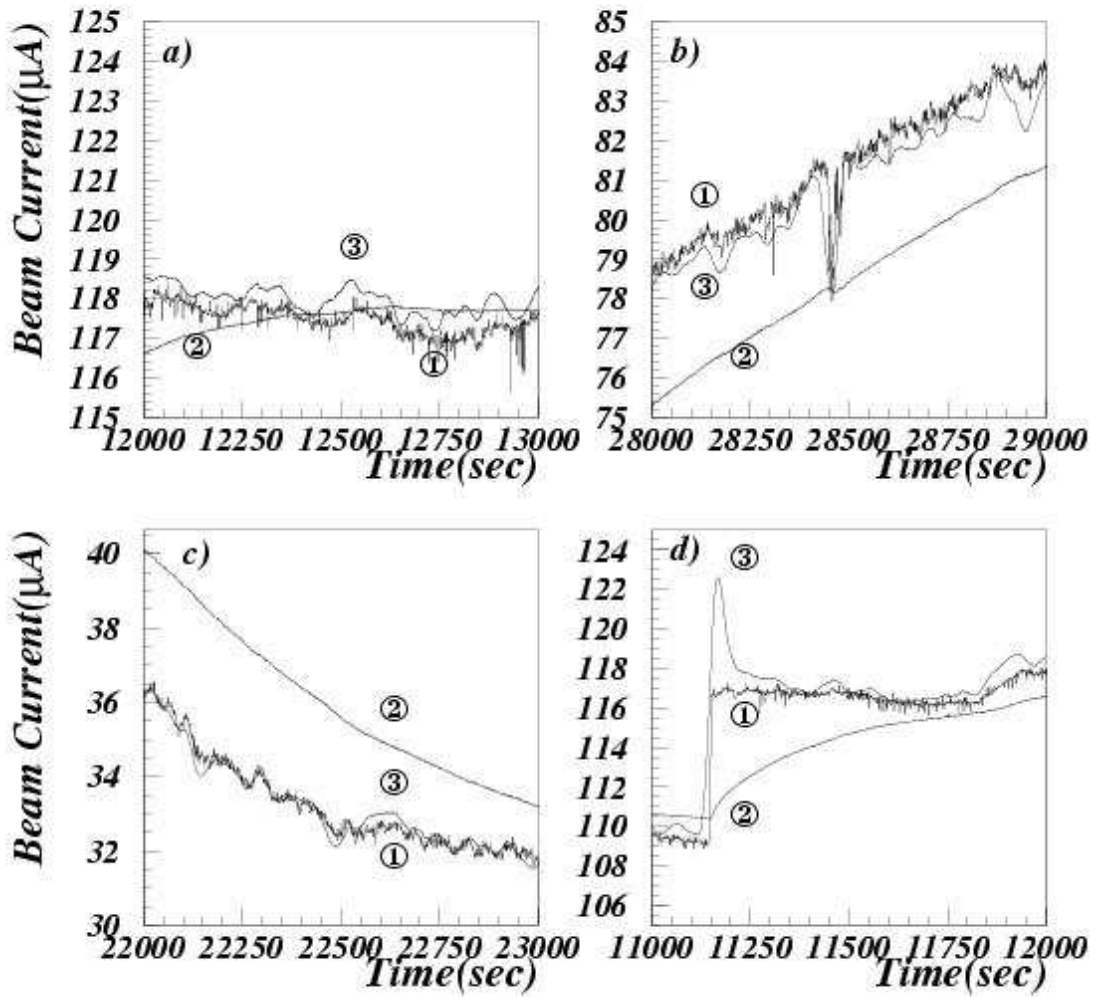}
\caption{
Beam current,
$\bigcirc \hspace*{-7pt} 1$ measured by the Faraday cup, 
$\bigcirc \hspace*{-7pt} 2$, calculated from  $k_1 \cdot H$ 
and $\bigcirc \hspace*{-7pt} 3$ 
$k_1 \cdot H+k_2 \cdot {\delta H}/{\delta t}$ 
as a function of time. 
The measurements  were carried out for 
a)stable beam current 
b)increasing beam current 
c)decreasing beam current 
and 
d)for rapid beam step of current. 
}
\label{fig:fandh}
\end{figure}

In order to determine the parameters $k_1$ and $k_2$, 
an experiment was carried out with a \nuc{3}{He}$^{2+}$ beam of 40 keV 
({
E$_{\rm cm}$}). 
The beam current was calculated from the HFS output by comparison 
with the Faraday cup measurements. 
The HFS output was measured by a KEYTHLEY 2000 multimeter. 
The parameters $k_1$ and $k_2$ were determined as follows: 
\\1) Parameter $k_1$ : If the system is stable, that is, 
incident beam $I$ and temperature $T$ are stable, 
the second term of Eq. (2) can be ignored. 
In this condition, the parameter $k_1$ can be determined 
by a least square fit as shown in Fig. \ref{fig:k2}(a). 
\\2) Parameter $k_2$ : This papameter can be determined 
once parameter $k_1$ is known, as seen in Eq. (2). 
The term $\delta H/\delta t$ was measured for averaged time scales of 
3, 7, 15, and 30 s. 
{
It was found that the scale of 30 s was appropriate. }
Fig. \ref{fig:k2}(b) shows $\delta H/\delta t$ as a function of $I - k_1 \cdot H$. 
\\3) Comparison with beam current: Fig. \ref{fig:fandh} shows the beam current 
as a function of time measured using a Faraday cup, 
$k_1 \cdot H$, and $k_1 \cdot H  + k_2 \frac{dT}{dt}$ 
with different beam conditions, i.e., 
stable, slowly increasing, and decreasing beam current. 

The calculated currents measured with the HFS  
reproduce the measurements with the Faraday cup as shown 
in Figs. \ref{fig:fandh}(a) to \ref{fig:fandh}(c). 
On the other hand, 
if the beam current is suddenly changed as shown in Fig. \ref{fig:fandh}(d), 
the calculated currents from the HFS output overestimate the value measured by the HFS. 
Further improvements of this system are necessary. 
Fortunately, this case should not be a fatal problem 
for an astrophysical experiment 
with long term measurement times of typically one day or one month, 
since this occurs rarely, 
at most one or two times in a day. 
Thus, in the experiment, the error caused by this overestimate can be neglected.

\subsubsection{Reproducibility of beam current for different energies}
\begin{figure}
\includegraphics*[width=150mm]{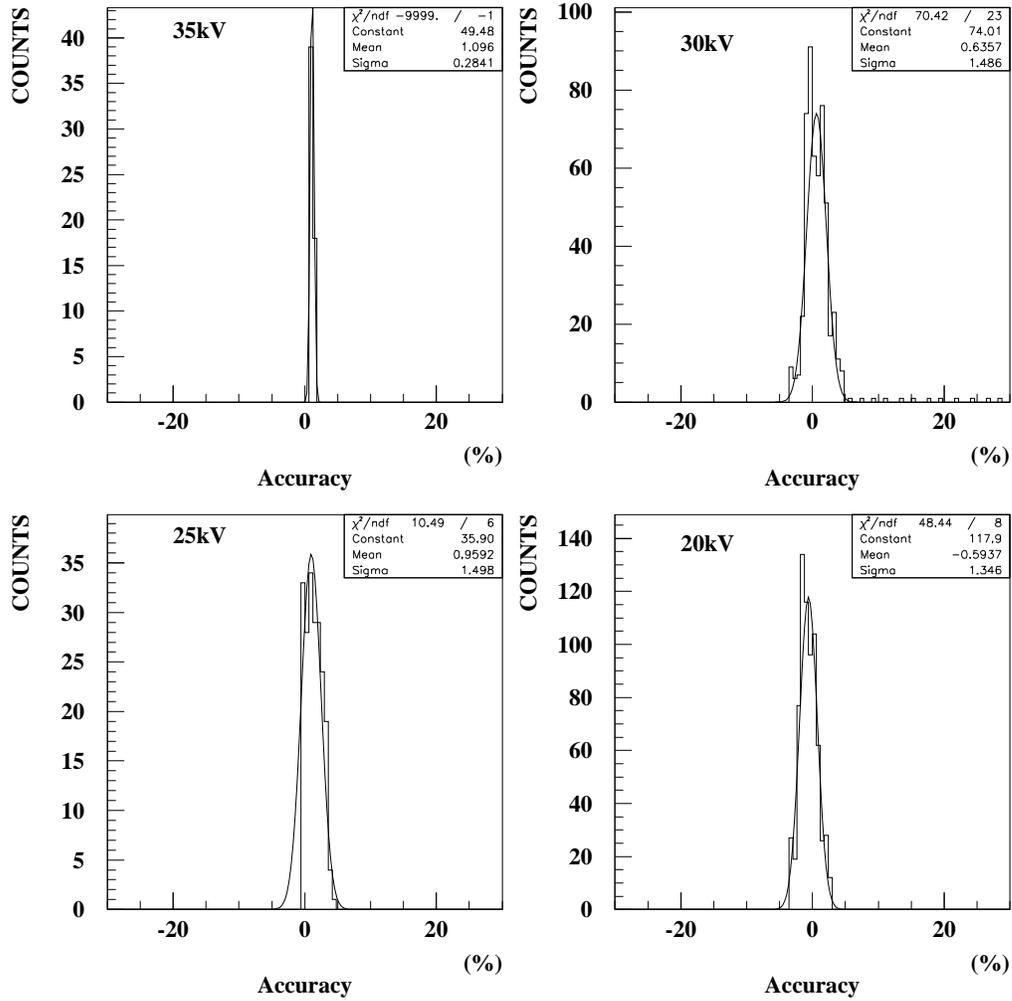}
\caption{
The accuracy of the beam current for different beam energies 
35, 30, 25, and 20 keV, in the form of (IHFS-IFC)/IFC, 
where IHFS and IFC denote the beam currents measured by the HFS and the Faraday cup(FC). 
}
\label{fig:accu-4}
\end{figure}

The reproducibility of the beam current determined from the HFS output 
was verified for several beam energies. 
Measurements were carried out 
{
on \nuc{3}{He}$^{2+}$ beam } 
at incident kinetic energies of 35, 30, 25, and 20 keV. 
The beam currents were calibrated using the parameters determined with the 40 keV beam. 
Fig. \ref{fig:accu-4} shows the accuracy of the calculated beam current 
in the form of $(I_{HFS}-I_{FC})/I_{FC}$, 
where $I_{HFS}$ and $I_{FC}$ are the beam currents 
measured by the HFS and the Faraday cup (FC), respectively. 
The accuracy was better than 2 \% for the measured energies. 
 
\subsubsection{Estimation for transferred heat in HFS}
The heat transferred through HFS can be calculated 
from the calibration parameter of HFS (OMEGA HFS3: 1.10 $\sim$ 1.11 $\mu$V/W/m$^{2}$ 
at 70 $^{\circ}$F). 
Also the kinetic energy of incident particles 
can be calculated from the charge integration. 
Table \ref{tab:heattrs} shows the results of these calculations. 
It was found that in the equilibrium state most of the heat 
is transferred through the HFS.

\subsubsection{Heat exchange with the surroundings}
Because of the different vacuum conditions 
during the calibration ($\sim$10$^{-6}$ Torr) and during the experiment (0.1 Torr), 
the effects of convection heat losses 
by the flowing gas in the target chamber have to be estimated. 
The heat transmission by convection was measured by comparing the transferred heat 
through the HFS for the two conditions of vacuum (10$^{-6}$ Torr
and 0.1 Torr) in the target chamber. 
The temperature of cool base was 30 $^{\circ}$C. 
The difference of HFS output was 0.04 W 
corresponding to about 1.7 \% for a 35 keV 100 $\mu$A beam. 


\subsection{dE-E counter telescope}
\begin{figure}[hbtp]
\includegraphics*[width=150mm]{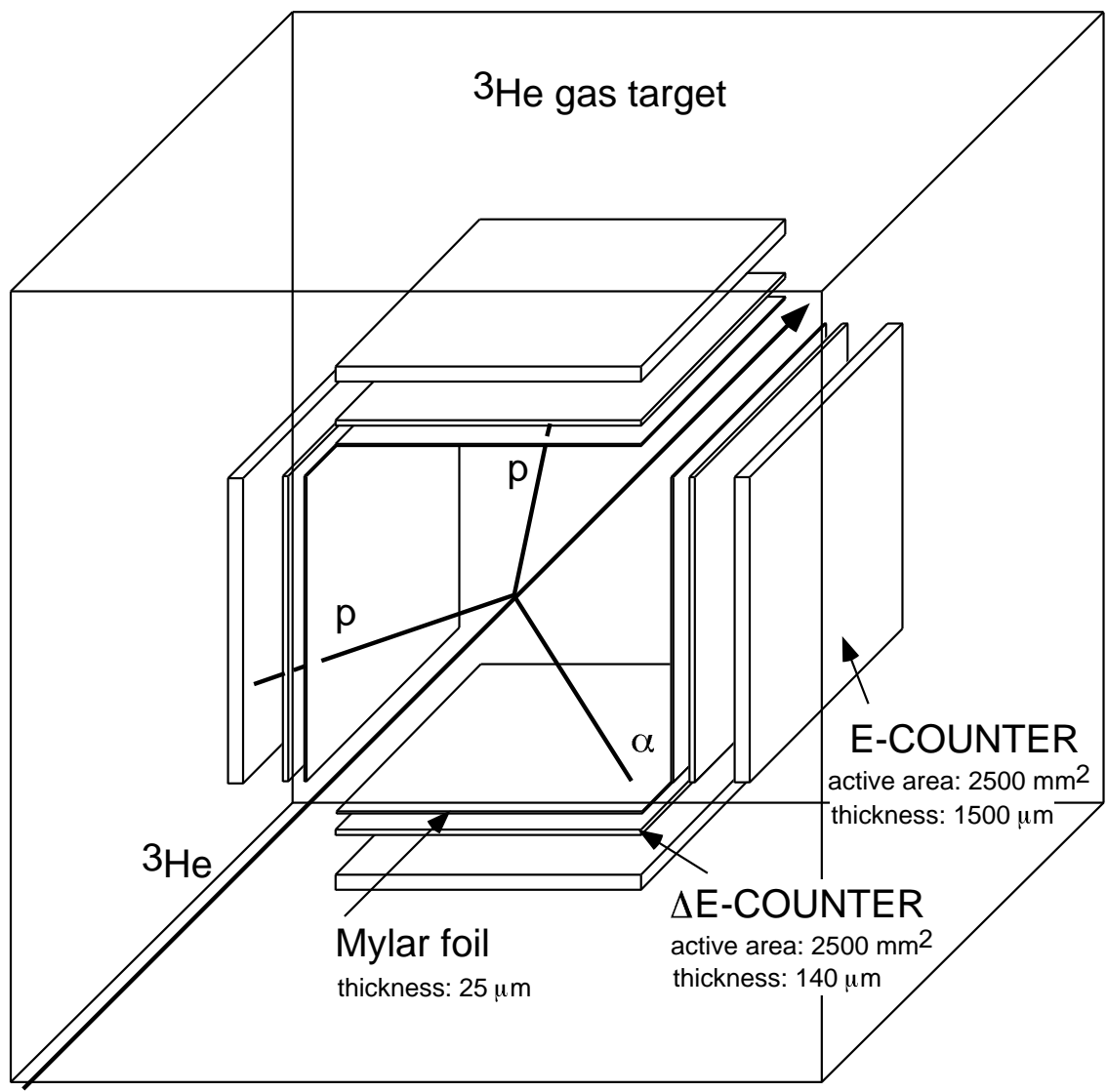}
\caption[detector]{
Schematic view of detector assembly. 
}
\label{fig:detector}
\end{figure}

In order to ensure a large detection efficiency 
and a clear discrimination of real events, 
we exploit four dE-E counter telescopes by using semiconductor detectors 
for the measurement of the \nuc{3}{He}(\nuc{3}{He}, 2p)\nuc{4}{He} reaction. 
These detectors are installed into the target chamber filled with \nuc{3}{He} gas 
and are capable of identifying the \nuc{3}{He}+\nuc{3}{He} reaction 
(Q = 12.86 MeV) as shown in Fig. \ref{fig:detector}. 

The reaction generates two protons which have kinetic energies of 0 to 10.7 MeV, 
and an alpha particle which has kinetic energies of 0 to 4.3 MeV. 
The dE and E detectors in each telescope have an active area 2500 mm$^2$, 
the dE detector has a thickness of 140 $\mu$m 
and the E detector has a thickness of 1500 $\mu$m (MICRON Ltd. ). 

To stop the generated alpha, 
photons and elastically scattered \nuc{3}{He} from the beam, 
an aluminized Mylar films 
with thickness of 25 $\mu$m are located in front of all dE counters. 
The distance between the dE-counter or E counter and the beam axis is 32.5 mm 
and 37.1 mm, respectively. 
These detectors are fixed to a cylindrical and four faced base of 
oxygen free high conductive copper (OFHC). 
It is helpful to avoid microphonic noise and natural background. 

\subsection{Data acquisition system}
\begin{figure}[hbtp]
\includegraphics*[width=150mm]{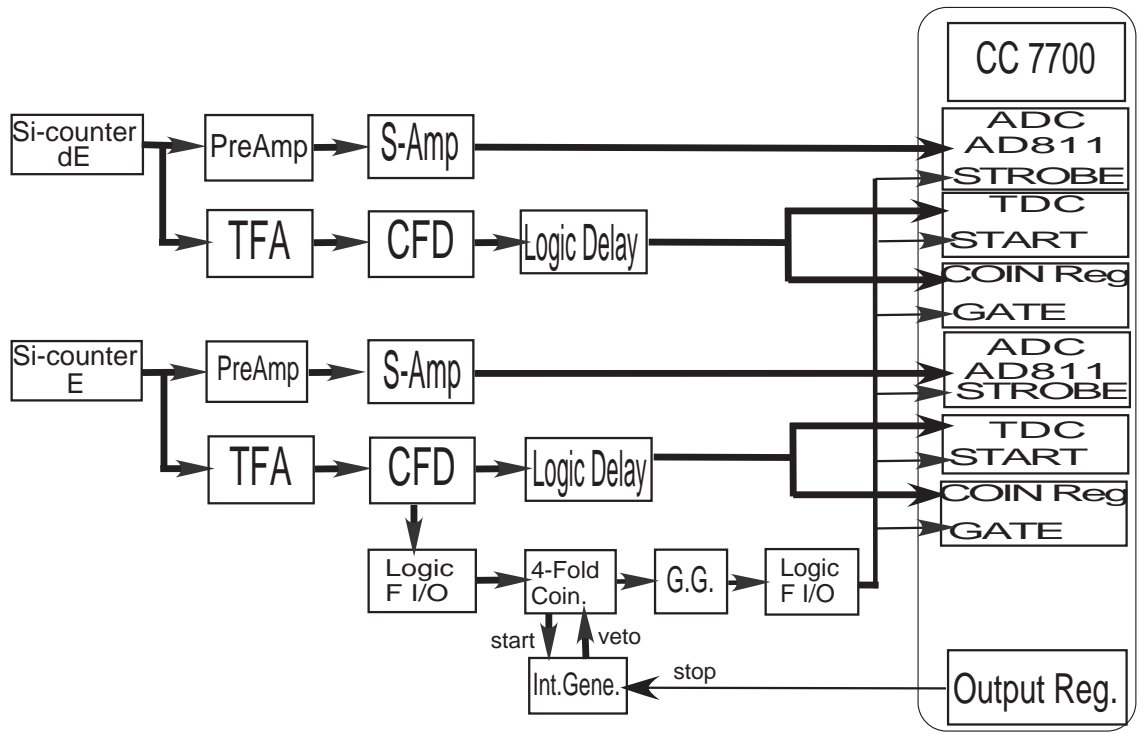}
\caption[detector]{
View of electronics for data acquisition. 
}
\label{fig:daq}
\end{figure}

Analog signals from each detector are fed into preamplifiers 
(model 142IH, ORTEC, for dE counters model 142B, ORTEC, for E-counter) 
with inorganic coaxial cables (Cu-S5ESS-05 DIPSOL CHEMICAL Co.,Ltd.).  
Since a shorter distance between the detector and preamplifier 
is desirable to reduce electrical noise, the distance is 45 cm. 
The signals from the preamplifiers are amplified by both spectroscopy amplifiers 
(SAMP, model 472 ORTEC) and timing filter amplifiers (TFA, model 454, ORTEC). 
The signals from the SAMP are led to a CAMAC peak 
sensitive analog-digital-converter system (ADC ,model AD811, ORTEC) 
while the signal from the TFA's are sent to a system of constant fraction discriminators 
(CFD, model 935, ORTEC) where the thresholds are set 
above the noise level of the detectors. 
The logic output of the CFD is fed into a logic Fan-In/Fan-Out 
(Fan-I/O, model 429A, ORTEC) delivering a gate signal for the CAMAC ADCs 
with a gate width of 400 nsec. 
The logic output provides a start signal for a time spectrum 
via a CAMAC time-digital-converter system (TDC, model 2228A, Lecroy). 
The stop signal of the TDC's is provided by the CFD 
with a 100 nsec delay. 
The signals from the ADC and TDC systems are contlolled 
by a CAMAC crate controller (model CC7700, TOYO). 
The data from the crate controller 
is transferred to Linux station (model L400c, DELL), via a CAMAC bus,
and stored on hard disk. 
A schematic diagram of the present data acquisition system 
is shown in Fig. \ref{fig:daq}. 

The dead time of this data taking system is 400 $\mu$sec for 1 event. 
The typical counting rate of the measurement of the \nuc{3}{He} + \nuc{3}{He} reaction, 
which includes the background events caused by \nuc{3}{He} + d reaction, 
cosmic ray, and electrical noise is usually about 3 counts/sec. 
Therefore, the total dead time of these measurements is about 0.1 \%. 

Before the reaction experiments, all the counters were calibrated 
using a \nuc{241}{Am} $\alpha$ source (5.48 MeV). 
The energy resolution of the dE-counters was 100 $\sim$ 120 keV (FWHM) 
for a 5.48 MeV $\alpha$ particle, 
and for the E-counter it was 70 $\sim$ 100 keV(FWHM). 
The energy gain of S-amp was optimized to be able to measure the energy range 
up to 10 MeV for the dE-counter and 20 MeV for the E-counter. 
This energy range is required to measure not only \nuc{3}{He} + \nuc{3}{He} events 
but also \nuc{3}{He} + d events, 
since \nuc{3}{He} + d events are needed for estimation of background 
and they are useful for checking the energy scale of each counter telescope. 
The linearity of the present amplifier system was measured 
by use of a precision pulse generator (model 419, ORTEC). 
Linearity was observed to within 0.05 \% 
for all of the energy range of each counter. 
The stability of the energy gain of the amplifiers 
was checked and the resultant gain shifts were less than 2 \% for 6 months. 
 
\section{Analysis of experiments}
For evaluation of the cross section, 
the number of counts for the \nuc{3}{He}(\nuc{3}{He},2p)\nuc{4}{He} reaction, 
the \nuc{3}{He} target density and the \nuc{3}{He} beam intensity 
should be measured. 
The values for the effective reaction energy and the integral term 
for detection efficiency were calculated by means of a Monte Carlo simulation 
developed in the frame of the present work. 

The number of counts $dN(z)$ per unit of time with respect to a length $dz$ 
of the extended \nuc{3}{He} gas target is given by the expression
\begin{equation}
dN(z)=N_{\rm t} \cdot N_{\rm b} \cdot \sigma(E(z))\cdot
\eta(z)\cdot dz,
\label{eq:loss1}
\end{equation}
where
$N(z)$ is the number of counts for the \nuc{3}{He}(\nuc{3}{He},2p)\nuc{4}{He} reaction, 
$N_{\rm t}$ is the \nuc{3}{He} target density, 
$N_{\rm b}$ is the \nuc{3}{He} beam intensity per unit time, and 
$\eta(z)$ is the absolute detection efficiency. 

Introducing the stopping power $\epsilon$  
(i.e. the energy loss per unit length), equation (\ref{eq:loss1}) can be rewritten in the form 
\begin{equation}
dN(E) = N_{\rm t} \cdot N_{\rm b} \cdot \sigma(E) \cdot
\eta(E) \cdot  \epsilon (E)^{-1} \cdot dE.
\label{eq:loss2}
\end{equation}
The total number of counts for the full target length is then given by 
\begin{equation}
N = N_{\rm t} \cdot N_{\rm b} \cdot \int _{L} \sigma(E) \cdot
\eta(E) \cdot  \epsilon (E)^{\rm -1} \cdot dE.
\label{eq:loss3}
\end{equation}

For the case of a thin target, introducing an effective reaction energy 
$E_{\rm eff}$ corresponding to the mean value of the projectile energy 
distribution in the detection setup, one arrives at 
\begin{equation}
 N = N_{\rm t} \cdot N_{\rm b} \cdot \sigma(E_{\rm eff}) \cdot 
  \int _{L} \eta(E) \cdot  \epsilon (E)^{\rm -1} \cdot dE. 
\label{eq:loss4}
\end{equation}

\subsection{Effective reaction energy}
The effective reaction energy was the mean value of the beam energy 
derived from an energy loss calculation in the target gas. 
As we could not apply any measurement for the absolute energy of ion beam, 
such as time of flight techniques 
or an Wien filter, 
we determined the value 
by measuring a voltage divided 
with a precise register chain of a ratio of 1/10000(STANDARD ENERGY, S-100) 
for the applied voltage to the ion source 
(SPELLMAN, SL-1200 (60 kV/20mA)). 
This resistive voltage divider was investigated 
by applying the exact voltage calibrated with the second standard 
and the resultant absolute accuracy is $\pm$0.2 \%. 
It was measured at intervals of 1.5 second for all measurements. 
The stability of the voltage was less than 0.1 \% for about one day. 

Compared with an ion source of a quiescent plasma 
such as a duoplasmatron ion source an ECR ion source has a finite plasma potential. 
Thus, we took into account this plasma potential for the acceleration voltage. 
The adopted value was 21.3$\pm$2.4 eV as reported by JAERI group 
for the NANOGUN ECR ion source \cite{Saitoh}. 

For the extended geometry in the present gas target experiment, 
the reaction energy distribution due to the energy loss of the ion beam 
along the beam path should be estimated precisely as possible. 
In low energy experiments this might raise the ambiguity 
for the electron screening potential; 
we had to take care of experimental conditions 
such as target pressure or its difference along the beam axis. 
In this series experiments for energies less than 30 keV center of mass energy, 
this problem should be treated more rigorously. 

There are quite number of experimental and theoretical works 
for stopping powers of charged particles in matter. 
Charged particles lose their energy through collisions 
with nuclei and with atomic electrons in matter. 
Although the greatest part of the energy loss occurs by collisions with electrons, 
low-energy ions lose their energy by collisions not only with electrons 
but with nuclei. 
Since it is impossible to deduce stopping powers data near the zero energy,
using present-day technology,  
we have to use updated compilations with an accuracy of ranges between 2 \% 
and 10 \% \cite{SP}. 
Therefore to calculate the energy loss in the target, 
we used the stopping power values estimated by the SRIM computer code \cite{SRIM}, 
which gives results which are consistent with the experimental energy losses 
to within a 10 \% difference at most.

\begin{figure}[hbtp]
\begin{center}
\includegraphics*[width=150mm]{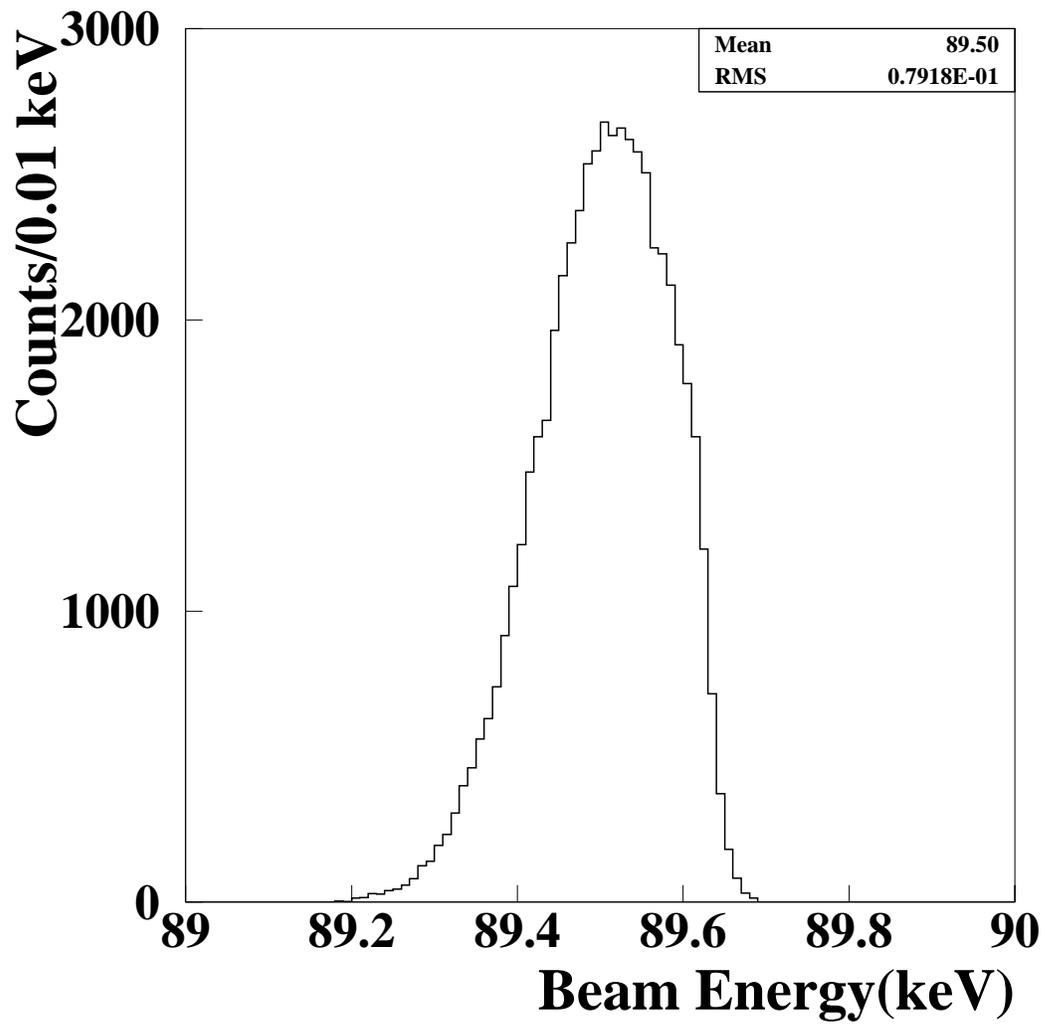}
\caption[dedx]{
Simulated interaction energy distribution for \nuc{3}{He}+\nuc{3}{He} reaction at
$E_{lab}$=90.0 keV. A target gas pressure of 0.1 Torr is assumed.
}
\label{fig:dedx}
\end{center}
\end{figure}

For example, 
the stopping power of incident \nuc{3}{He} with an energy of E$_{\rm lab}$=90.00 keV 
is 9.3$\times$10$^{-15}$ eV/atom/cm$^{2}$. 
The energy distribution of this particle 
in the \nuc{3}{He} gas target 
with pressure of 0.1 Torr is simulated by the Mote Carlo program 
and the result is shown in Fig. \ref{fig:dedx}.

We employ the full target length of $L$ =30 cm as the distance 
between the entrance of the target just after the collimator 
and the entrance of the beam calorimeter. 
Since the rapid reduction of the cross section 
is about 11.2 \% at E$_{\rm LAB}$(\nuc{3}{He})=90 keV over the target thickness 
when we assumed a constant $S(E)$ factor, 
the effective energy loss is evaluated to be (500$\pm$50) eV. 
This value roughly agrees with the back-of-the-envelope value of 493 eV 
for the energy loss between the entrance and the center of the 
counter telescope on the beam axis. 
As shown in Fig. \ref{fig:dedx}, 
we simulated the effective target length and estimated energy spread of 79 eV 
as the error of the incident beam energy. 
The energy loss due to the residual gas between the ion source and the target entrance 
is 3.7$\times$10$^{-3}$ \% of incident beam energy. 

In summary, at an incident beam energy of (90.0$\pm$0.13) keV, 
the effective reaction energy is E$_{\rm lab}$ = (89.50$\pm$0.13 ) keV, 
taking into account the accuracy of 0.1 \% for the acceleration voltage, 
10 \% for the stopping power and 0.09 \% for the energy spread in the target. 

\subsection{Beam current}
The incident projectile number ($N_b$) is deduced from the deposited power 
measured with our calorimetric device as described 
in the former section. 
It was calibrated using a charged beam in vacuum by comparing  
with the electrical beam current in the Faraday cup. 
The electrical charge collected in the Faraday cup 
was measured with the current integrator (KEYTHLEY 616 digital electrometer). 
The absolute value of the current integrator was calibrated by measuring the current 
which was supplied with a precise current source (R6161 ADVANTEST). 
It has an accuracy and a stability better than 0.001 \%. 
The difference between current measured by current integrator 
and the value of the current source is less than 3 \% and it was corrected. 
{
The following was taken into account to evaluate the incident particle number $N_{b}$. 
The beam current measured by calorimeter was corrected 
because of the energy loss of incident beam in the target gas. 
The energy loss was estimated by SRIM program and was 1.60$\pm$0.16 keV 
when incident beam energy was 90.0 keV (E$_{\rm CM}$=45.0 keV). 
This energy loss was 1.8 \% lower than the incident energy. }

\begin{figure}[hbtp]
\begin{center}
\includegraphics*[width=150mm]{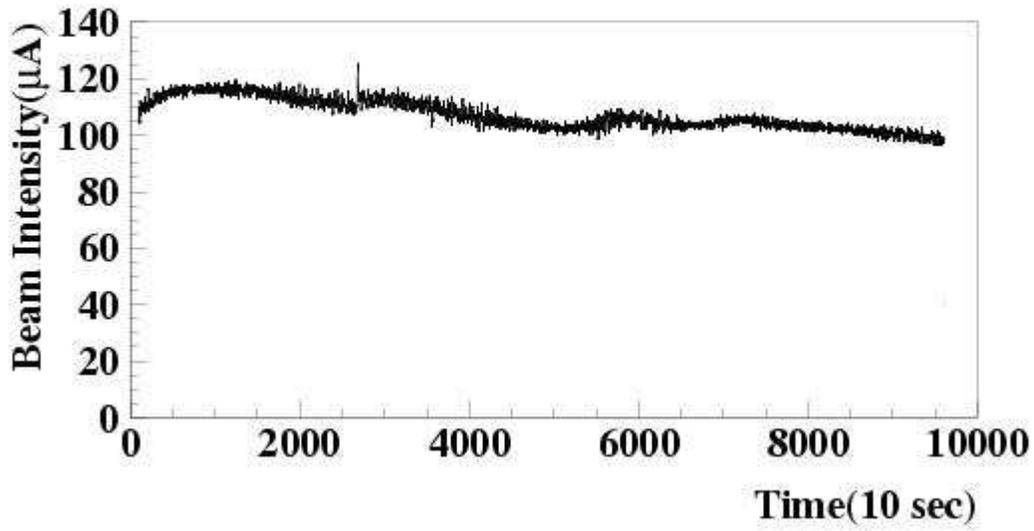}
\caption[beam-curr]{
Measured and corrected intensity of \nuc{3}{He}$^{2+}$ beam 
at $E_{\rm lab}$=90.0 keV. 
Measurements were made at 1.5 second intervals. 
}
\label{fig:beam-curr}
\end{center}
\end{figure}

The intensity was simultaneously corrected by recording the beam energy 
and the target gas pressure at intervals of 1.5 seconds. 
The typical beam intensity was about 100 $\mu$A at an incident energy of 90 keV 
for the \nuc{3}{He}$^{2+}$ beam. 
The beam intensity measured by the present system during experiment 
is shown in Fig. \ref{fig:beam-curr}.

\subsection{Target density}
There are several factors which affect the target density $N_t$ 
such as the gas temperature and a pressure gradient in the target chamber. 
We measured the target gas temperature 
with a termister (103AT-2) inside the chamber 
which was likely to be different from that of the laboratory room 
since the target gas was heated by the beam and cooled by the circulated gas 
for the purification system. 

\begin{figure}[hbtp]
\begin{center}
\includegraphics*[width=150mm]{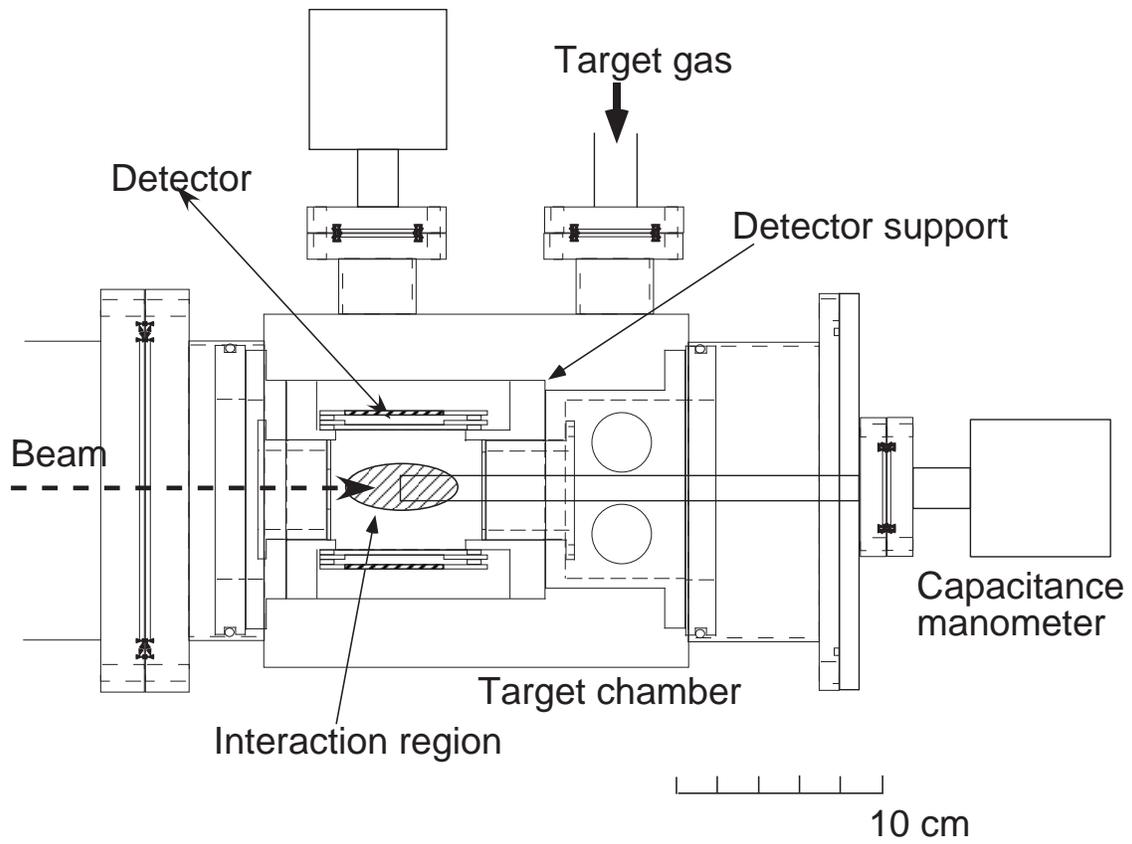}
\caption[target-chamber]{
Schematic view of the measurement of the gas pressure at the interaction
region between the beam and gas target.
}
\label{fig:target-chamber}
\end{center}
\end{figure}

\begin{figure}[hbtp]
\begin{center}
\includegraphics*[width=150mm]{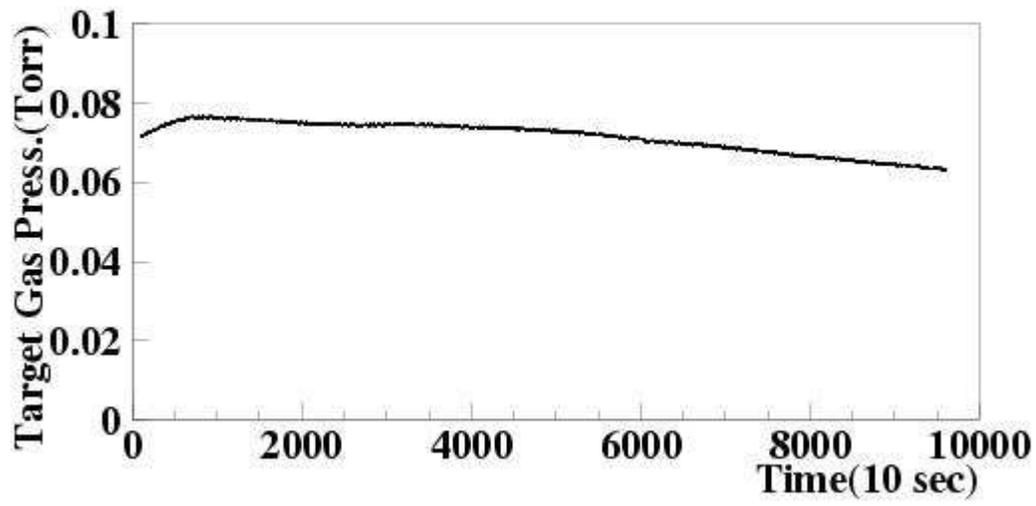}
\caption[target-press]{
Measured target \nuc{3}{He} gas pressure by Baratron capacitance manometer. 
Measurement intervals were 1.5 sec for all experiments for 
the \nuc{3}{He}+\nuc{3}{He} reaction. 
The pressure was normalized to that of temperature at 0 $^{\circ}$C. 
}
\label{fig:target-press}
\end{center}
\end{figure}

As the target pressure could not be measured directly at the beam-target interaction region  
during the experiment, 
the pressure was measured at the top of the target chamber 
as shown in Fig. \ref{fig:target-chamber}. 
The pressure distribution caused by the geometry of detector holder, 
collimators and gas circulation 
was measured by extending the stainless tube 
directly from the capacitance manometer set downstream of the target chamber. 
The capacitance manometer, which is usually installed at the top of the target chamber, 
was removed to the end of the chamber only at this measurement, 
as shown in Fig. \ref{fig:target-chamber}. 
Simultaneously, we used another gauge just 
before the inlet to the chamber for a normalization. 
The difference between the target gas pressure at the top of the chamber 
and that measured at the interaction region was rather small, 
which might be  a shorter mean-free path 
at gas pressure of the order of 
{
0.1} Torr.  
The absolute pressure at the target should be corrected by 5 \% 
less than that measured at the top of the chamber. 
Owing to these corrections, 
the target density can be determined to an accuracy of 1.3 \%, 
considering the accuracy of 0.16 \% from the target gas pressure, 
1 \% from the correction due to the gas temperature, and 0.8 \% from the correction 
due to the measurement position. 
The measured and corrected target gas pressure at the every interval of 1.5 seconds 
is shown in Fig. \ref{fig:target-press}.

\section{Data analysis}
           
\subsection{Monte Carlo simulation for the OCEAN experiment}
In order to find an optimum detector setup 
for a high efficiency and background free measurement, 
we exploit the Monte Carlo simulation program based on GEANT3. 
It was used to calculate the interaction between the ejectiles and the detectors. 
Also, the GENBOD code was used to generate the ejectiles. 
Thus the program takes into account the following aspects: 
1) the detector geometry, 
2) the energy loss and energy straggling of the ejectiles 
in both the target gas and i the thin foil in front of the detector, 
3) kinematic effects on the energy of the ejectiles in the target, 
4) yield dependence of the ejectiles over the passage of the target, and 
5) the non-uniform depletion thickness for the E counter. 
 
\subsection{Measurement of the D(\nuc{3}{He},p)\nuc{4}{He} reaction}
To verify the validity of the simulation program, that is 
to estimate the systematic error, 
the experimental results of the D(\nuc{3}{He}, p)\nuc{4}{He} reaction (Q=18.4 MeV) 
are compared with those of simulations. 
The comparison to this reaction has several advantages: 
1) The generated protons from the \nuc{3}{He} + d reaction have a definite 
energy of 14.7 MeV. 
2) The energy of the protons from this reaction 
is almost the same as that from the \nuc{3}{He} + \nuc{3}{He} reaction. 
3) the cross section of the \nuc{3}{He} + d reaction is six orders of magnitude 
larger than that of \nuc{3}{He} + \nuc{3}{He} reaction.

The D(\nuc{3}{He},p)\nuc{4}{He} reaction was performed by 
using 90 keV (E$_{\rm cm}$=45 keV)\nuc{3}{He}{2+} beam at the OCEAN facility. 
The target pressure of the deuterium gas 
was maintained around 1.0$\times$10$^{-4}$ Torr. 

\begin{figure}[hbtp]
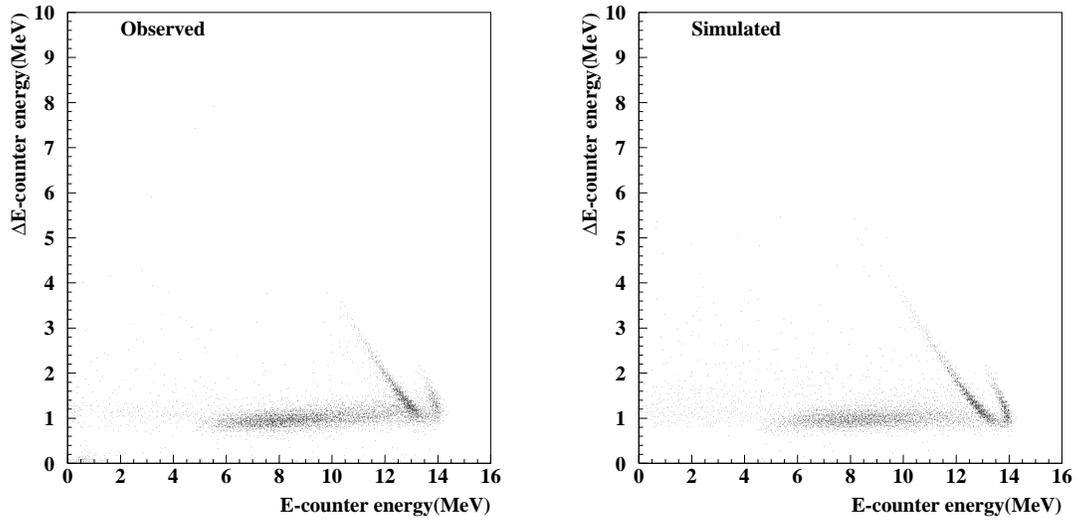

\begin{center}
\includegraphics*[width=75mm]{hed-dee-exp.epsf}
\includegraphics*[width=75mm]{hed-dee-monte.epsf}
\caption{
The observed (left) and simulated (right) energy spectrum for the d($^{3}$He,p)$^{4}$He 
reaction at E$_{^{3}He}$=90.0 keV 
by means of the $\Delta$E-E counter telescope. 
}
\label{fig:hed-dee-exp}
\end{center}
\end{figure}

Fig. \ref{fig:hed-dee-exp} shows observed and simulated energy spectra 
obtained with the dE-E are compared.  
The broad energy spectra for E = 5$\sim$14 MeV at $\delta$E = 1 MeV  
arise from an insufficient depletion depth for protons 
incident on the surface at angle near 90 degree. 
We applied a bias voltage of 180 V to the E-counter, 
which corresponds to the depletion depth of 900 $\mu$m, 
to avoid a discharge in the gas target thicker depth needs more voltage.  
The other two structures arise from kinematics effects 
in combination with protons which are incident 
at angular ranges of 135$^{\circ}$$\sim$180$^{\circ}$ 
and 0$^{\circ}$$\sim$45$^{\circ}$. 
These three remarkable features are well simulated in the energy spectrum 
as shown in Fig.  \ref{fig:hed-dee-exp} (b).

\subsection{Background analysis}
\begin{figure}[htb]
\includegraphics*[width=150mm]{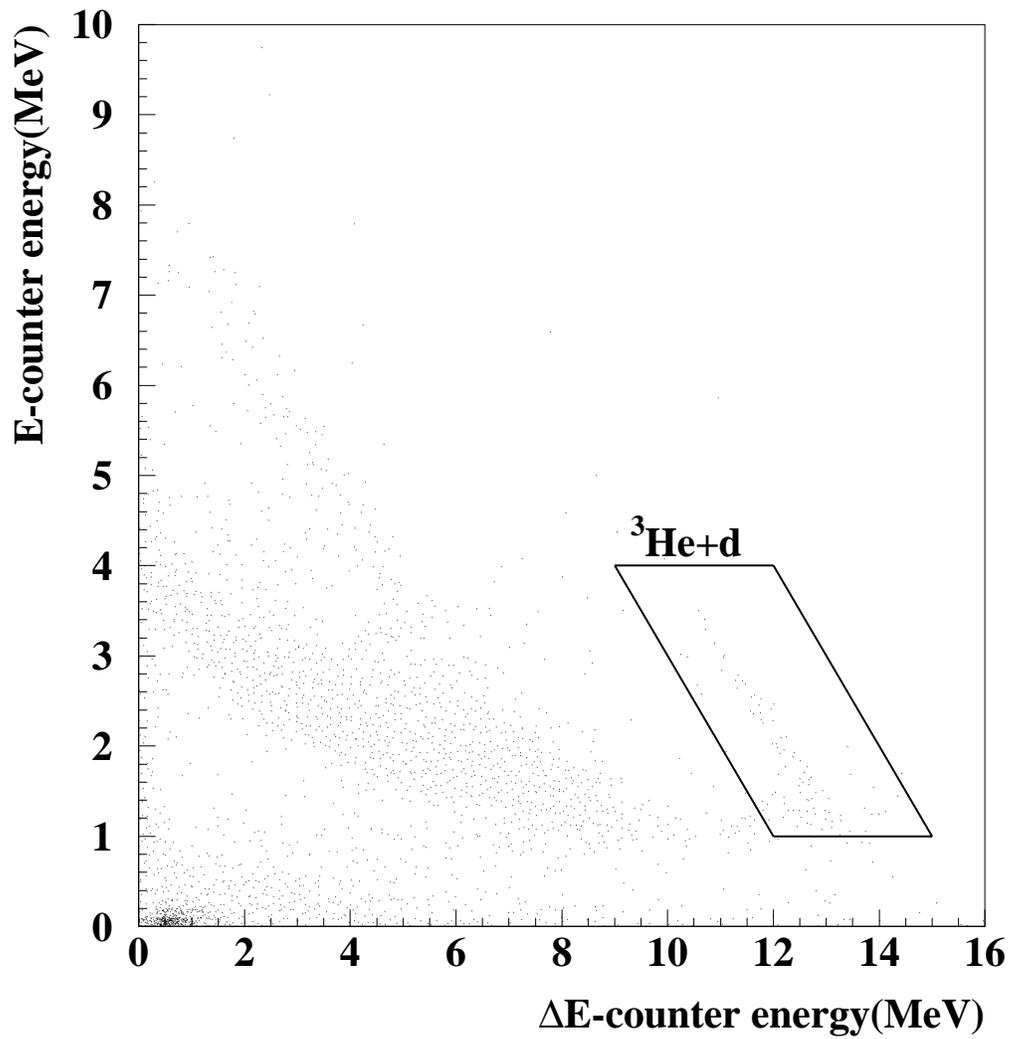}
\caption[rough-cut]{
Observed $\Delta$E-E energy spectrum of the 
\nuc{3}{He}(\nuc{3}{He},2p)\nuc{4}{He} and 
D(\nuc{3}{He},p)\nuc{4}{He} reactions. 
The solid line shows the acceptable energy region 
for the D(\nuc{3}{He},p)\nuc{4}{He} reaction. 
The deuteron contamination in the target \nuc{3}{He} gas 
was evaluated from the events in this region. 
}
\label{fig:rough-cut}
\end{figure}

It is crucial for the present measurement of the 
\nuc{3}{He}(\nuc{3}{He},2p)\nuc{4}{He} reaction
to identify the background origin and to discriminate the true events 
from the fake events. 
As already stated, deuterium contamination in the target are 
the most serious. 
The number of deuterons in the gas target was determined 
from the data during the measurement of 
the \nuc{3}{He}(\nuc{3}{He},2p)\nuc{4}{He} reaction 
as shown in Fig. \ref{fig:rough-cut}. 
For this estimate, the value of the cross section 
for the D(\nuc{3}{He},p)\nuc{4}{He} reaction 
was taken from ref. \cite{HED}. 
We conclude that the deuterium contamination is 0.2 ppm in the target gas, 
and that such a level could make background events 
at least 0.1 \% of the observed events of the \nuc{3}{He} 
(\nuc{3}{He},2p)\nuc{4}{He} reaction at the energy of E$_{\rm cm}$=45 keV. 
 
\begin{figure}[hbt]
\begin{center}
\includegraphics*[width=150mm]{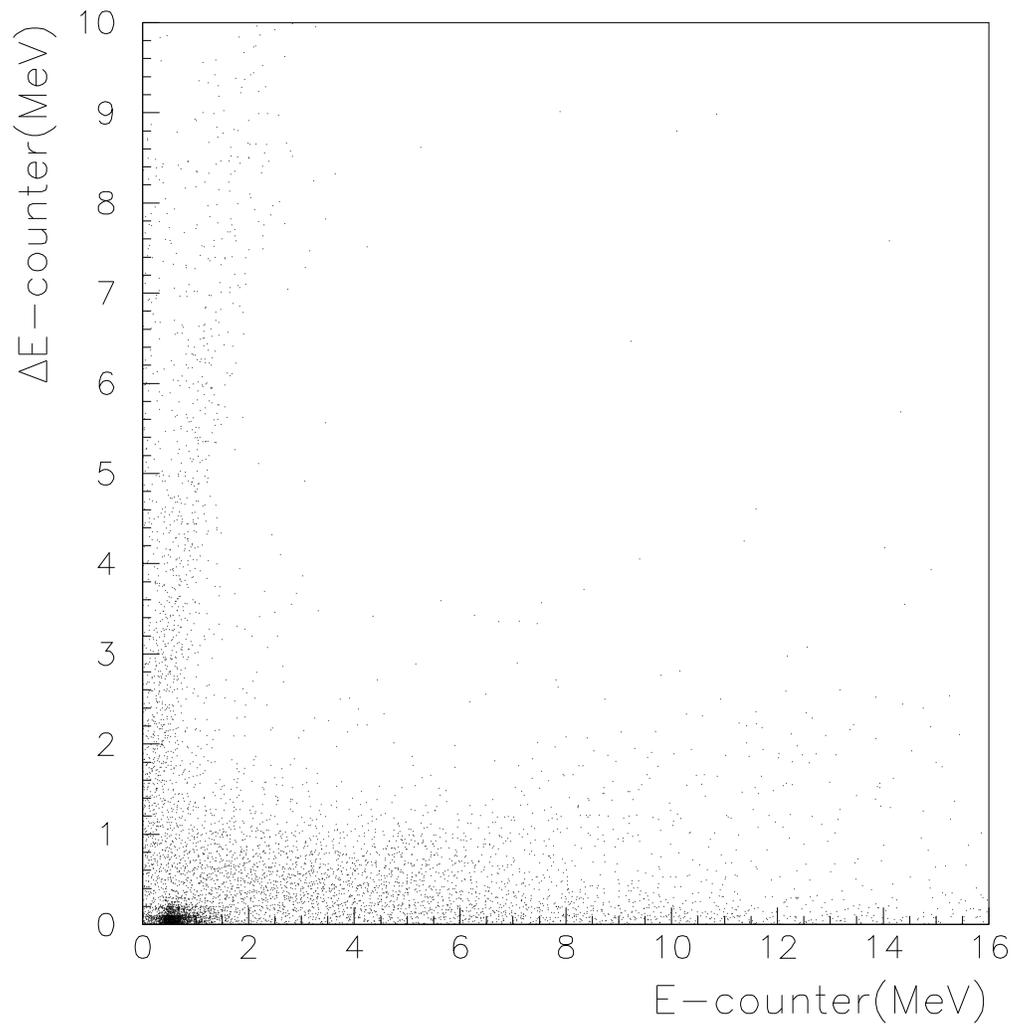}
\caption[bg]{
Background events arising from electronic noise and cosmic-ray,
observed with the same geometry 
as the \nuc{3}{He}(\nuc{3}{He},2p)\nuc{4}{He} experiment.
}
\label{fig:bg}
\end{center}
\end{figure}

Another source of background events arises from electrical noise 
and cosmic rays. 
These are observed during the measurement 
without \nuc{3}{He} beam for 38 days of operation of OCEAN 
as shown in Fig. \ref{fig:bg}. 
The contribution from this background to the window 
of the \nuc{3}{He}(\nuc{3}{He},2p)$\alpha$ reaction is 3.6 counts/day. 
Of these, the cosmic muon events are located around dE = 70 keV and E =450 keV 
because of the minimum ionization loss of 2 MeV cm$^2$/g. 
We attempted to reject these events by applying the veto-counter upper and lower places 
for the target chamber.  
Finally the expected rate of the present reaction 
at lower energy is around a few events per day or less, 
and a typical single background rate of silicon detectors 
is one event per hour or more. 
In order to remove such accidental events, 
two proton coincidence should inevitably be required for the identification 
of the present reaction near the Gamow peak (next series of OCEAN experiment). 

\subsection{Detector efficiency}
\begin{figure}[hbt]
\begin{center}
\includegraphics*[width=150mm]{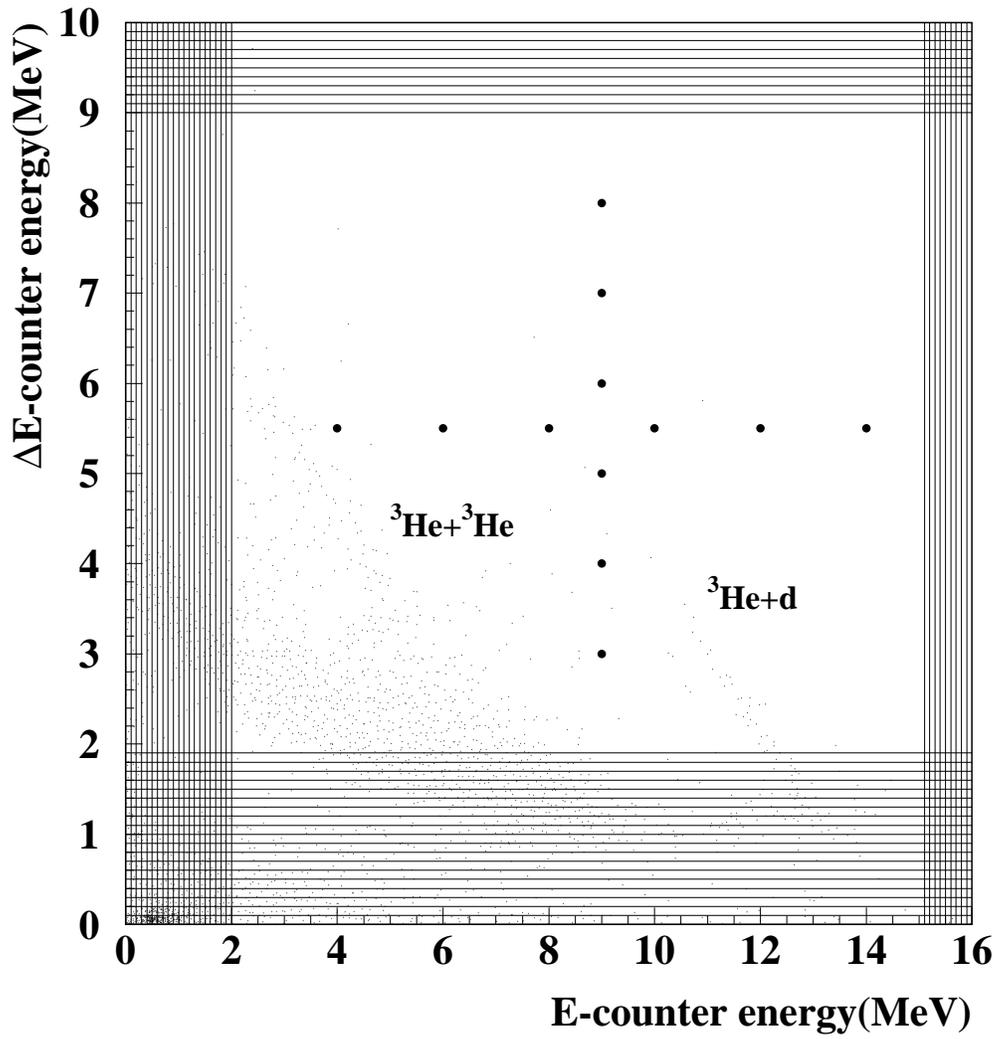}
\caption[cut-point]{
Schematic view of the the acceptance region.
The $\Delta$E-E energy scatter region was divided into
16000 parts of 100 keV $\times$100 keV divisions. 
Signal to noise ratio was examined in each part. 
}
\label{fig:parts}
\end{center}
\end{figure}

\begin{figure}[hbt]
\begin{center}
\includegraphics*[width=150mm]{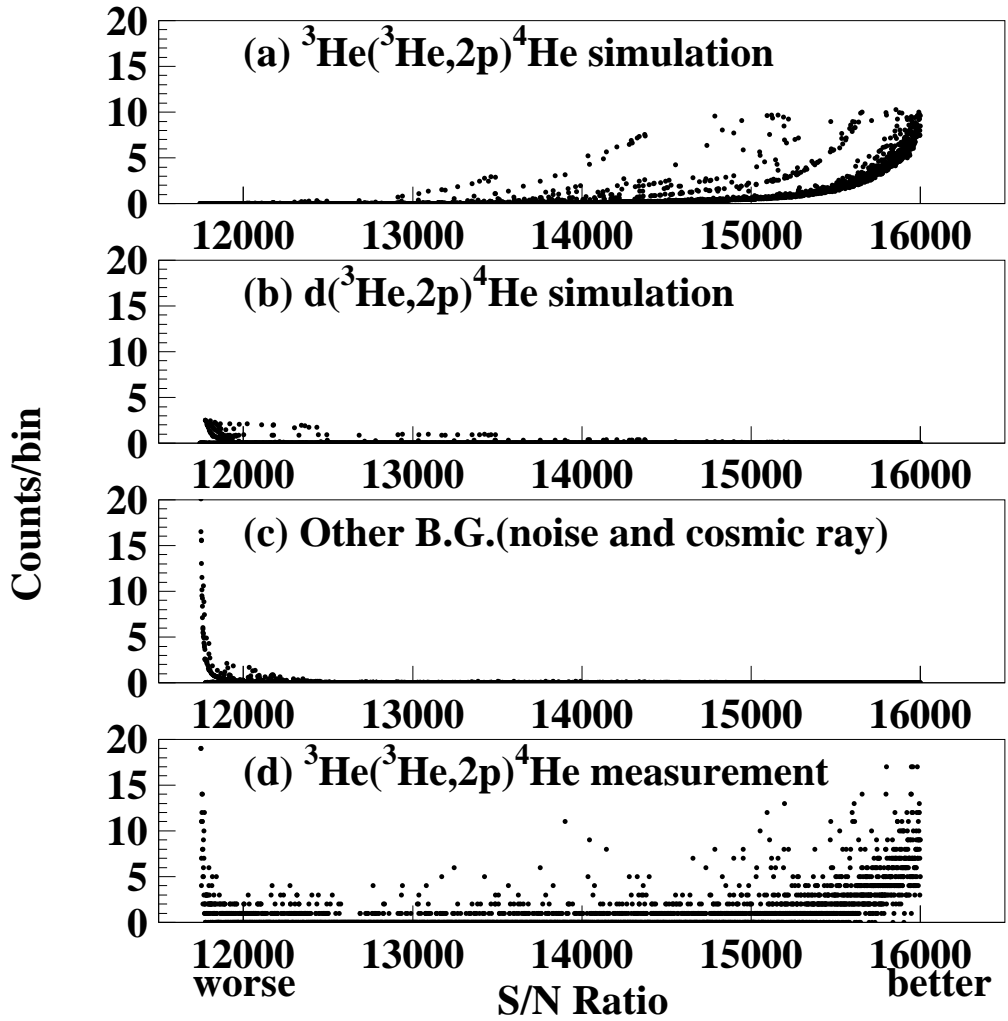}
\caption[cut-point]{
Event distribution ordered as a function of their S/N ratios for four types of data. 
The parts having a better S/N locate at the right hand side while worse parts are
located to the left. 
(a)simulated $^{3}$He+$^{3}$He, (b)simulated $^{3}$He+d, 
(c)other observed background(electric noise and cosmic-ray), 
(d)observed $^{3}$He+$^{3}$He are shown. 
S/N is given by S(simulated $^{3}$He+$^{3}$He)/ 
N(simulated $^{3}$He+d + observed other background). 
}
\label{fig:cut-point}
\end{center}
\end{figure}

We developed a reasonable method to determine the acceptable area 
for the real events of the \nuc{3}{He}(\nuc{3}{He},2p)\nuc{4}{He} reaction 
in the dE-E scatter plot  
without a redundant and ambiguous procedure. 
Four types of data such as observed events of the reaction, 
simulated events for the \nuc{3}{He}(\nuc{3}{He},2p)\nuc{4}{He} reaction 
and for the D(\nuc{3}{He},p)\nuc{4}{He} reaction, 
and observed background events 
are summarized for the analysis of each experimental run. 
The energy distribution of the dE-E scatter plot is divided into 16000 parts 
of 100 keV $\times$ 100 keV parts as shown in Fig. \ref{fig:parts}. 
The signal to noise ratio which should be derived from the Monte Carlo 
MC ( \nuc{3}{He} + \nuc{3}{He} ) divided by MC (\nuc{3}{He} + D) 
and  measured background  
was allotted for 16000 parts. 
All parts are ordered as a function of their S/N; 
parts having a better S/N are located at the right hand side 
while worse parts are located to the left,  
as shown in Fig. \ref{fig:cut-point}. 
Fig \ref{fig:cut-point}(a) shows the distribution of the simulated events from 
the \nuc{3}{He} + \nuc{3}{He} reaction as a function of S/N ratio. 
Many events are located in the right hand side in the figure, 
that should correspond to the better S/N. 
Also, Figs. \ref{fig:cut-point}(b) and (c) show 
the distribution of the simulated 
events for \nuc{3}{He} + d reaction events 
and the observed background events as discussed above. 
It is usual that these background events should be located 
in the left hand side of each figure. 
Fig. \ref{fig:cut-point}(d) shows the distribution of the observed event for 
\nuc{3}{He} + \nuc{3}{He} reaction 
at E$_{\rm cm}$=45 keV, as a function of S/N. 
The contribution from the background events is apparently very small. 
Therefore the observed distribution as shown in Fig. \ref{fig:cut-point}(d) is 
particularly similar to the simulated one, 
shownn in Fig. \ref{fig:cut-point}(a), 
without subtraction of any background events 
as shown in Fig. \ref{fig:cut-point}(d).  
    
\begin{figure}[hb]
\begin{center}
\includegraphics*[width=150mm]{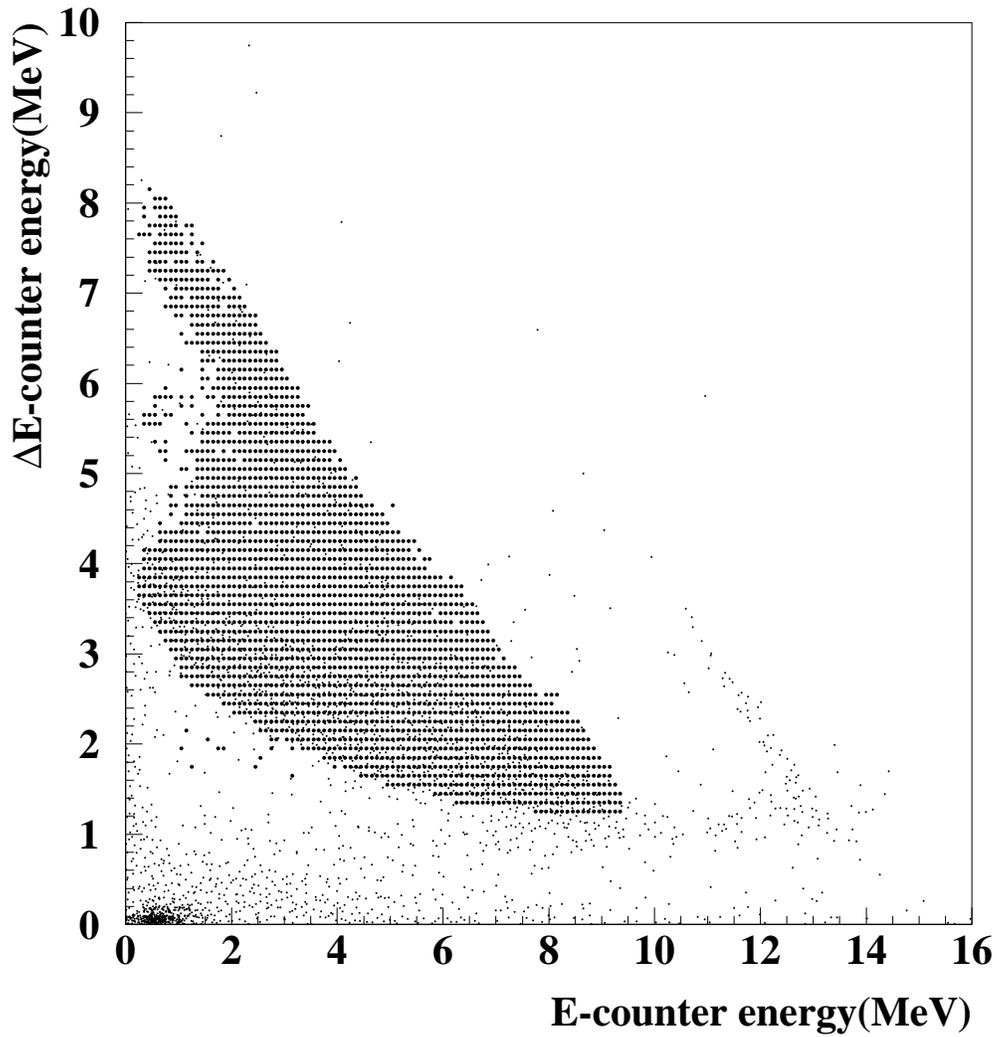}
\caption[hehe-area]{
$\Delta$E-E scatter plot obtained 
from the $^{3}$He+$^{3}$He reaction experiment.
The region assembly of grosspoints shows the accepted region as the true events 
from the $^{3}$He($^{3}$He,2p)$^{4}$He reaction. 
}
\label{fig:hehe-area}
\end{center}
\end{figure}

Since, most of the background events exist in channels less than 13000 
(Fig. \ref{fig:cut-point})
the acceptable area for the \nuc{3}{He} + \nuc{3}{He} reaction 
could be assigned to channels larger than 13000 channel. 
The region of that is shown in Fig. \ref{fig:hehe-area}. 

\begin{figure}[hbt]
\begin{center}
\includegraphics*[width=150mm]{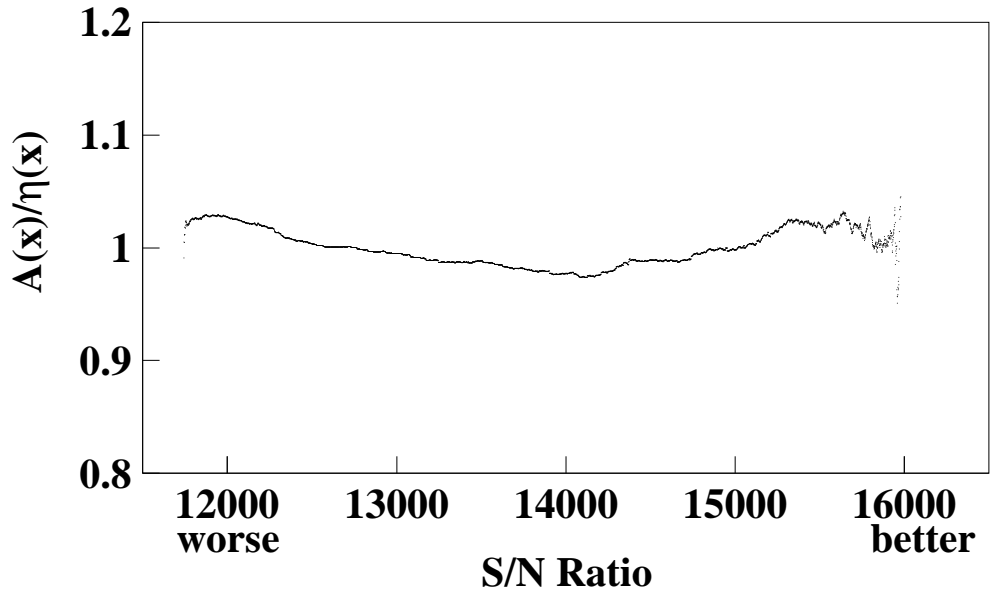}
\caption[kaitest-accu]{
Ratio of $A(x)/ \eta(x)$ as the function of the boundary parameter $x$. 
The ratio corresponds to the cross section of the $^{3}$He+$^{3}$He reaction.
$\eta(x)$ and $A(x)$ was derived from eqs. (\ref{eq:he3-simulated}) and 
(\ref{eq:he3-observed}) [see text], respectively.
}
\label{fig:kaitest-accu}
\end{center}
\end{figure}

Experimental results are as follows; 
3344 counts are observed in the acceptable region, 
while the contribution from the \nuc{3}{He} + D events to the region is 20.9 counts, 
and that from the other background component is 2.46 counts. 
After subtracting the number of these background events 
from the number of observed events in the acceptable region, 
the number of true events for the \nuc{3}{He} + \nuc{3}{He} reaction is 3337.4, 
with a statistical error of 1.8 \%. 
According to this procedure, the detection efficiency $\eta(x)$ can be written as 
\begin{equation}
\eta(x)=\sum_{i=x}^{16000}N_{\rm a}(i), 
\label{eq:he3-simulated}
\end{equation}
where $i=x$ to 16000, 
$N_{\rm a}(i)$ is the number of counts for the simulated distribution, 
and $x$ is the parameter of the boundary cut point for the accepted events. 
The accepted event have been derived as 
\begin{equation}
A(x)=\sum_{i=x}^{16000}\{ N_{\rm d}(i)-N_{\rm b}(i)-N_{\rm c}(i)\}, 
\label{eq:he3-observed}
\end{equation}
where, 
$A(x)$ refers to accepted events 
for the \nuc{3}{He}(\nuc{3}{He},2p)\nuc{4}{He} reaction, 
$N_d (i)$, $N_b (i)$, $N_c (i)$ are the 
number of counts for the observed events, 
for the d(\nuc{3}{He},p)\nuc{4}{He} 
and for other background events, respectively. 
Finally the ratio of $A(x)/y(x)$ corresponding to the cross section 
of the \nuc{3}{He}(\nuc{3}{He},2p)\nuc{4}{He} reaction can be obtained. 
The ratio slightly depends on the boundary parameter $x$ 
as shown in Fig. \ref{fig:kaitest-accu}. 
Thus, the accuracy of the simulated energy distribution in a scatter plot 
should be derived from the fluctuation of this ratio. 
When we include the geometrical uncertainty of the counter telescope, 
i.e., 0.5 \% uncertainty of the detection efficiency 
simulated with Monte Carlo program, 
the systematical error of the detection efficiency is evaluated to be 3 \%. 

\section{Experimental results and discussion}
The experimental results for the cross section and S-factors 
obtained for the first series of OCEAN experiments as from the year 2000, 
together with the experimental conditions such as the live time, 
beam current, the target gas pressure, 
and the target temperature are shown in table \ref{tab:result}. 
The observed events for the \nuc{3}{He}(\nuc{3}{He},2p)\nuc{4}{He} 
reaction and the background events from various sources 
are also shown in the same table \ref{tab:result}.

The cross section for the \nuc{3}{He}(\nuc{3}{He},2p)\nuc{4}{He} reaction 
has been derived from the following equation, 
\begin{equation}
 N = N_{\rm t} \cdot N_{\rm b} \cdot \sigma(E_{\rm eff}) \cdot
  \int _{L} \eta(E) \cdot  \epsilon (E)^{\rm -1} \cdot dE.
 \label{eq:loss5}
\end{equation}
where $N$ is the number of counts for \nuc{3}{He}(\nuc{3}{He},2p)\nuc{4}{He} reaction, 
$N_{\rm t}$ is the \nuc{3}{He} target density, 
$N_{\rm b}$ is the \nuc{3}{He} beam intensity, 
$E_{\rm eff}$ is the effective reaction energy, 
$\eta(E)$ is the absolute detection efficiency, 
and $\epsilon (E)$ is the stopping power. 

The astrophysical S-factors were deduced from the equation 
\begin{equation}
\sigma(E)=\frac{S(E)}{E}\exp(-2\pi \eta),
\label{eq:s-factor}
\end{equation}
where $\eta$ is the Sommerfeld parameter given by: 
\begin{equation}
2\pi \eta =31.29Z_{1}Z_{2}(\frac{\mu}{E})^{1/2}
\label{eq:sommer}
\end{equation}
where
$Z_1$ and $Z_2$ are the nuclear charges of the interacting particles 
in the entrance channel, 
$\mu$ is reduced mass (in units of amu) 
and $E$ is the center of mass energy (in units of keV). 

Our present data between E$_{\rm cm}$= 45 keV to 31 keV  of S(E) are in good agreement 
with the results of existing data of Krauss et al \cite{LUNA2}.  
The accuracy of both statistical and the systematical data 
of the present measurement is better than that in ref. \cite{LUNA2}. 

For the last two decades, 
studies of the \nuc{3}{He}(\nuc{3}{He},2p)\nuc{4}{He} reaction 
have been carried out over a wide range of energies 
(Bacher and Tombrello \cite{Bacher}, 
Dwarakanath and Winkler \cite{Dwarakanath},   
Krauss et al \cite{LUNA2}, 
and LUNA \cite{LUNA}). Of the previous studies, 
experiments between 17.9 to 342.5 keV center of mass energy by 
Krauss et al cover a wider energy range than others 
since they used two accelerators of 350 kV  accelerator 
in Munster University and 100 kV facility at Bochum University. 
Recently the LUNA group in the LNGS has presented data down to 16.50 keV 
center of mass energy, 
although the data was obtained with two separate experimental set-ups,namely 
at the 450 kV accelerator at Bochum and also 50 kV at LNGS. 
There remains unaccesible region of energy between 45.82 keV and 24.80 keV. 
The data in this energy region has been supplied by Krauss et al. 

In these previous experiments there were many 
efforts to obtain the true scarce events 
from the contaminated background events, 
such as deuterium contamination both in the incident beam and in the target gas, 
cosmic rays (mainly muon) or heavy particles, and electric noise. 
There have been several solutions for these difficulties; 
Krauss et al., pointed out in 1987 the purity of the ion beam 
and of the target were of special interest and they estimated that at 350 keV 
the mass-3 beam contamination HD$^{+}$ was of the order of 10$^{-5}$ \cite{LUNA2}. 
Also they applied proton-proton coincidences to discriminate the real events 
from intruded events for the D(\nuc{3}{He},p)\nuc{4}{He} reaction. 
They surrounded the target chamber with NE102A plastic scintillator 
in order to identify for cosmic events. 
In this way the unidentified cosmic background coincidence event rate 
was estimated to be less than 1 event/200h in the measurements at E$_{\rm cm}$ $<$ 50 keV. 
Another source of background events due to occasional discharges 
of the accelerator high voltage could be avoided by amplifier filter or noise filter. 
In spite of these efforts, they still consider the contribution of the events 
caused by the high energy protons from the \nuc{3}{He}(d,p)\nuc{4}{He} reaction 
(e.g. 2.5 \% contribution at E$_{\rm cm}$=25 keV ). 
They also evaluated that the background contribution to the \nuc{3}{He}+\nuc{3}{He} region 
is 0.40 \% of the observed counts of the d-\nuc{3}{He} events. 
 

The measurements at the DTL showed that cosmic rays produced events 
within the dE-E region of the \nuc{3}{He}(\nuc{3}{He},2p)\nuc{4}{He} reaction 
amounts to 3.5$\times$10$^{-4}$ events/s, 
while in the latter place at LNGS this rate was observed to be reduced 
by at least a factor 200, 
that is, 1.8$\times$10$^{-6}$. 
It will be negligibly small for the low energy measurement 
less than 30 keV center of mass energy. 

We now argue that the OCEAN facility could overcome 
these difficulties by applying the following, 
while there still exist the possibilities of background 
due to the target gas: 

1) Only OCEAN exploits doubly charged He-3 ions produced with an ECR ion source 
and it could avoid the background events due to D (beam) + \nuc{3}{He}. 
Krauss et al performed a p-p coincidence experiment 
to discriminate the true event, 
but it could not be applied for the center of mass energy less than 25 keV. 

2) Our facility OCEAN has been located in an experimental area of the cyclotron 
and shielded with 5 m of thick concrete. 
We have observed the background events arising from cosmic rays in 38 days 
and also observed the effect from the induced events 
due to the high energy particles by using Veto-counters 
composed of plastic scintillators located up and down the target chamber. 
By this condition of the OCEAN sight we still have a sufficient signal to noise ratio 
of 10 times larger than unknown fake events as listed in table \ref{tab:result}. 

Generally, there are still many discussions 
for the screening potential 
which enhance the cross section remarkably for low energy fusion reactions. 
A recent result of the \nuc{3}{He}(\nuc{3}{He},2p)\nuc{4}{He} reaction data 
assumed for electron screening with U$_{\rm e}$ = 330 eV 
and adopted a polynomial function for $S(E)$ factor,namely 
$S(E) = 5.18 + 2.22 E + 0.804 E^2 $. 
The $S(0)$ value is close to the recommended value of AD98 (5.4 + 0.4 MeV b). 
In a review article concerning synthesis of the elements in stars \cite{REVM}, 
it cited LUNA data as 
$S(E_0) = 5.3 \pm 0.05(stat.) \pm 0.30(syst.) + 0.30(U_e) $ MeV b.  
It still includes a $U_e$ value which suggests 
that the effective value of $U_e$ may be larger than the adiabatic limit. 
 
Furthermore, it is mentioned that 
improved data should be obtained at energies from E = 25 keV to 60 keV 
to confirm or reject the possibility of a relatively larger systematic error 
in the $S(E)$ data 
around these energies.

\section{Conclusion and perspectives}
The present experiment with OCEAN has proved to reinforce the compilation data for nuclear 
astrophysics in solar fusion rates. 
High current and low background system could measure the cross 
section in the center of mass energy range from 50 to 30 keV. 
As a second series of experiments, 
we have started to measure at energies less than 30 keV, and these
results will be reported in the near future. 
Better results with respect to reduced stematic and statistical errors 
compared to existing data are expected. 
When we apply a beam intensity of 1 mA for the \nuc{3}{He}$^{2+}$ beam 
at E$_{\rm cm}$ = 20 to 30 keV 
and a target gas pressure of 0.1 Torr, 
we can expect 70 $\sim$ 2 events per day for real events, 
while the fake event will be 6 $\sim$ 1.8 events per day. 
Hence, we expect a measurement with a statistical error 
of about 10 \% in this energy range.

\begin{acknowledgments}
The authors gratefully acknowledge Professor Claus Rolfs for the design of the 
OCEAN accelerator facility. 
We acknowledge Professor T. Hasegawa for introducing ECR ion source 
as a bright \nuc{3}{He} particle source. 
We thank Dr. Uwe Greife and Dr. Mathias Junker 
for many suggestions to measure the reaction cross section of rare events 
from their experience. 
We thank Prof. H. Ejiri, Prof. Y. Nagai and Prof. H. Toki for 
various suggestions and discussions. 
We also thank Prof. G. Hillhouse for reading critical manuscript. 
This work was supported by the Grant-in-Aid of Scientific Research, 
Ministry of Education, Science, 
Culture and Sports, No. 08404015, and No. 0041120.
\end{acknowledgments}


\clearpage

\begin{table}
\caption{
The achieved beam current at the ion source and the target.
}
\label{tab:is2target}
\begin{center}
\begin{tabular}{cccc}\hline
H.V.   &       & Target($\mu$A) & Ion source($\mu$A) \\ \hline\hline
40 kV   & 1$^+$ & 1208           & 3010               \\
        & 2$^+$ & 103            & 3000               \\
30 kV   & 1$^+$ & 1200           & 3800               \\
        & 2$^+$ & 35             & 1900               \\ \hline
\end{tabular}
\end{center}
\end{table}

\clearpage
\begin{table}
\caption{
Comparison of the heat calculated from the kinetic energy of incident particles and 
from the HFS output. 
The beam current is measured by means of a Faraday cup, 
and the Voltage at the IS(Ion source) determines the kinetic energy of 
incident particles. 
The beam power is calculated by multiplying 
these factors. 
The heat through the HFS is calculated using the calibration parameter, 
which is given as 1.10$\sim$1.11 $\mu$V/W/m$^2$@70$^{\circ}$F. 
}
\label{tab:heattrs}
\begin{tabular}{ccccc}\\ \hline
Beam Current & Voltage at IS & HFS Output & Energy of Beam & Heat Transfer\\
 ($\mu$A)    & (kV)          & (mV)       & (W)            & (W)  \\ \hline
 0.          & 0.            & -0.83      & 0.             & 0.17 \\
 96.4        & 40.0          & 20.86      & 3.84           & 3.75 \\
 69.7        & 20.0          & 7.65       & 1.39           & 1.37 \\
 74.3        & 25.0          & 9.88       & 1.86           & 1.78 \\
 82.5        & 35.0          & 15.66      & 2.89           & 2.81 \\ \hline
\end{tabular}
\end{table}

\clearpage
\begin{small}
\begin{table}
\caption{\small 
Summary of measurements for the \nuc{3}{He}(\nuc{3}{He},2p)\nuc{4}{He} 
reaction. 
L.T. : Live Time,  
B.C. : Beam Current, 
T.P. : Target Pressure,  
T.T. : Target Temperature, 
C.S. : Cross Section, 
True : $^3$He+$^3$He, 
BG1  :  BG $^3$He+d, 
BG2  :  BG Other, 
Cnt  : Counts, 
S-Fac: Astrophysical S-factor. 
}
\label{tab:result}
\vspace*{-5mm}
\centering
\begin{tabular}{p{1.3cm}p{1.3cm}p{1.3cm}p{1.8cm}p{1.3cm}p{1.3cm}p{1.3cm}p{1.3cm}p{2.cm}p{2.0cm}} \\ \hline
E$_{\rm{cm}}$  & L.T.  & B.C.  & T.P.  & 
T.T.  & True   & BG1   & BG2  &
C.S.  & S-Fac  \\ 
 (keV) & (sec) &  ($\mu$A) & (Torr) & 
 ($^{\circ}$C) &  (Cnt)  & (Cnt)  & (Cnt) &
 (barn) & (MeV$\cdot$b) \\ \hline
45.3 & 92567 &  104. & 7.46$\times 10^{-2}$ & 27.1 & 3276 & 20.9 & 2.46  & 1.53$\times 10^{-8}$ & 
 5.39$\pm$0.09 \\
43.3 & 78647 &  91.4 & 6.72$\times 10^{-2}$ & 27.3 & 1374 & 7.50  & 2.09  & 9.55$\times 10^{-9}$ & 
 5.43$\pm$0.14  \\
41.3 & 80687 &  100. & 6.74$\times 10^{-2}$ & 27.1 & 939  & 7.08  & 2.15  & 5.79$\times 10^{-9}$ & 
 5.51$\pm$0.18 \\
39.3 & 83109 &  87.4 & 7.26$\times 10^{-2}$ & 27.0 & 542  & 6.08  & 2.21  & 3.44$\times 10^{-9}$ & 
 5.69$\pm$0.25 \\
37.3 & 155442 & 112. & 8.24$\times 10^{-2}$ & 29.3 & 770  & 17.0 & 4.14  & 1.83$\times 10^{-9}$ & 
  5.46$\pm$0.20 \\ 
35.2 & 338862 & 100. & 8.21$\times 10^{-2}$ & 29.3 & 770  & 21.4 & 9.02  & 9.46$\times 10^{-10}$ &
 5.62$\pm$0.21 \\
33.1 & 615814 & 103. & 8.25$\times 10^{-2}$ & 30.4 & 691  & 11.4 & 16.4 & 4.52$\times 10^{-10}$ &
 5.48$\pm$0.22 \\
31.2 & 528134 & 93.6 & 8.28$\times 10^{-2}$ & 30.3 & 293  & 5.02 & 14.1 & 2.46$\times 10^{-10}$ &
 6.40$\pm$0.39 \\ \hline
\end{tabular}
\end{table}
\end{small}

\clearpage

\clearpage

\clearpage

\clearpage

\clearpage

\clearpage

\clearpage

\clearpage

\clearpage

\clearpage

\clearpage

\clearpage

\clearpage

\clearpage

\clearpage

\clearpage

\clearpage

\clearpage

\clearpage

\clearpage

\end{document}